\documentclass[journal,twoside,web]{ieeecolor}
\usepackage{jsen}
\usepackage{cite}
\usepackage{amsmath,amssymb,amsfonts}
\usepackage{graphicx}
\usepackage{textcomp}
\usepackage{eso-pic}
\usepackage{url}
\usepackage{fancyhdr}
\usepackage{tikz}
\usepackage{wrapfig}
\usepackage{algorithm}
\usepackage{algpseudocode}
\usepackage{subcaption}
\usepackage{booktabs}
\usepackage{threeparttable}
\def\BibTeX{{\rm B\kern-.05em{\sc i\kern-.025em b}\kern-.08em
    T\kern-.1667em\lower.7ex\hbox{E}\kern-.125emX}}
\definecolor{abstractbg}{rgb}{0.89804,0.94510,0.83137}
\begin{document}
\title{Physics-Guided Concentration Inference from Resistance Transients in a Mixed-Phase SnO-SnO$_2$ Carbon Monoxide Sensor with p-n Switching}
\author{Sani Biswas,  Preetam Singh, and Amit Kumar Gangwar
\thanks{Received 10 May 2026; revised 1 August 2026; accepted 6 August
2026.
S. Biswas acknowledges partial funding from Centro de Modelamiento Matem\'atico (CMM) FB210005 BASAL fund for centers of excellence  from ANID-Chile. A.K Gangwar acknowledges the ANID-Milenium Science Initiative Program (NCN20223\_07) for supporting this research work.
}
\thanks{S. Biswas is with the Centro de Modelamiento Matem\'atico, Universidad de Chile \& IRL 2807 - CNRS, Santiago,  Chile. (e-mail: sani.dumkal@gmail.com; sbiswas@dim.uchile.cl). }
\thanks{A. K. Gangwar is with the Department of Chemical Engineering, Biotechnology and Materials, FCFM, Universidad de Chile, Santiago,  Chile and ANID - Millenium Science Initiative, Millenium Nuclei of Advanced MXenes for Sustainable Applications (AMXSA), Santiago, Chile (e-mail: amitptl195@gmail.com; amit.gangwar@uchile.cl).}
\thanks{P. Singh is with CSIR-National Physical Laboratory, Dr. K.S. Krishnan Marg, New Delhi, 110012, India.  (e-mail:  singhp.nplindia@csir.res.in).}
\thanks{The authors gratefully acknowledge the prior experimental study reported in \cite{Gangwar2024},
which provided the experimental foundation for the present computational analysis.}
\thanks{Digital Object Identifier 10.1109/JSEN.2026.3723678}}

\IEEEtitleabstractindextext{%
\fcolorbox{abstractbg}{abstractbg}{%
\begin{minipage}{\textwidth}%
%\begin{wrapfigure}[12]{r}{3in}%
%\includegraphics[width=3in]{jsenga.png}%
%\end{wrapfigure}%
\begin{abstract}
This work presents a physics-guided machine-learning framework for carbon monoxide concentration
inference from experimentally measured resistance transients of a mixed-phase SnO-SnO$_2$ martials gas
sensor exhibiting temperature-dependent p-n switching behavior. Cycle-level transient responses are
represented through physically interpretable descriptors and complemented by compact fast Fourier transform (FFT)- and
discrete wavelet transform (DWT)-based summaries. Using leakage-aware grouped cross-validation, we study both multi-class
concentration classification and continuous concentration regression for the p-type and n-type
sensing regimes separately. Across both regimes, fused features provide the strongest overall
performance, while the physics-guided descriptor block remains highly competitive, indicating that
the dominant concentration information is already encoded in physically meaningful transient
dynamics. The p-type branch shows the best concentration-class discrimination, with the fused Random
Forest classifier reaching approximately $96.5\%$ accuracy, whereas the n-type branch yields the
best quantitative concentration estimation, with the fused Random Forest regressor achieving
$\mathrm{MAE}\approx 1.48\,\mathrm{ppm}$ and $R^2\approx 0.992$. These results reveal a clear
dual-regime behavior: p-type sensing is particularly favorable for classification, whereas n-type
sensing is more favorable for high-fidelity regression. More broadly, the study demonstrates that
leakage-aware, cycle-level, physics-guided machine learning can extend conventional gas-sensing
analysis beyond single-response metrics while preserving physical interpretability.
\end{abstract}

\end{minipage}}}

\maketitle

\AddToShipoutPictureFG*{%
  \AtPageLowerLeft{%
    \raisebox{0.30in}{%
      \hspace*{0.35in}%
      \parbox{7.15in}{%
        \centering
        \fontsize{7}{8}\selectfont
        1558-1748 \copyright\ 2026 IEEE. All rights reserved, including rights for text and data mining, and training of artificial intelligence\\
        and similar technologies. Personal use is permitted, but republication/redistribution requires IEEE permission.\\
        See \url{https://www.ieee.org/publications/rights/index.html} for more information.
      }%
    }%
  }%
}

\section{Introduction}
\label{sec:introduction}
% =========================================================
Carbon monoxide (CO) is a highly hazardous toxic gas whose reliable detection is essential for environmental monitoring, industrial safety, and domestic protection systems \cite{Mahajan2020}. Metal-oxide-semiconductor (MOS) gas sensors have attracted considerable attention because of their structural simplicity, low fabrication cost, and high sensitivity \cite{He2024}. Among them, mixed-phase SnO-SnO$_2$ sensors are attractive because the coexistence of p-type SnO and n-type SnO$_2$ enables temperature-dependent p--n switching, providing two complementary sensing regimes within a single sensing platform~\cite{Saruhan2021,Gangwar2024}.

Recent studies have demonstrated that machine-learning methods can improve gas discrimination and concentration estimation from transient responses in chemiresistive gas sensors. Random forest, support vector machines, and multi-layer perceptrons have been successfully applied to gas identification and concentration prediction, while fast Fourier transform (FFT)- and
discrete wavelet transform (DWT)-based representations have been explored to characterize transient-response dynamics for learning-based inference \cite{Acharyya2020,Acharyya2022a,Oh2022}. However, existing studies have not considered leakage-aware cycle-level learning for mixed-phase SnO-SnO$_2$ sensors with temperature-dependent p--n switching, nor have they systematically compared the p-type and n-type sensing regimes under a grouped-validation framework.

The present study builds upon the experimentally validated mixed-phase SnO-SnO$_2$ thin-film CO sensor reported in \cite{Gangwar2024}, which exhibits temperature-dependent p--n switching. The low-temperature branch operates in the p-type regime (approximately $150\,^{\circ}$C), whereas the high-temperature branch operates in the n-type regime (approximately $325\,^{\circ}$C), providing two complementary sensing modes within the same material platform. The measured resistance transients capture adsorption, saturation, and recovery dynamics across these sensing regimes.
While conventional gas-sensing analysis typically represents each sensing event using scalar quantities such as sensor response, response time, and recovery time, these metrics do not fully exploit the temporal structure of the resistance transient. Transients with similar peak responses may still exhibit distinct dynamic characteristics, indicating that additional concentration-related information is embedded in the transient shape beyond conventional scalar summaries~\cite{Clifford1982}.

The main contributions of this work are as follows:

\begin{enumerate}
\renewcommand{\labelenumi}{(\roman{enumi})}
\item formulation of an experimentally grounded cycle-level learning framework for CO sensing using mixed-phase SnO-SnO$_2$ resistance transients;

\item development of a physics-guided feature representation with complementary FFT- and DWT-based descriptors;

\item implementation of grouped classification and regression analysis to control correlation-induced data leakage;

\item separate quantitative assessment of the p-type and n-type sensing regimes, followed by a comparative interpretation of their concentration-inference behavior.
\end{enumerate}

The remainder of the paper is organized as follows. Section~\ref{sec:physical_background} presents the physical background and
sensing mechanism of the mixed-phase SnO-SnO$_2$ platform, together with the experimental basis
for the p-type and n-type regimes. Section~\ref{sec:experimental_data} describes the experimental data representation and the
cycle-level dataset construction. Section~\ref{sec:physics-guided} introduces the physics-guided preprocessing and
feature-design methodology, including the auxiliary FFT and DWT summaries. Section~\ref{sec:supervised_framework} presents the
machine-learning framework and validation protocol. Section~\ref{sec:result} reports the p-type and n-type results
for both concentration classification and concentration regression, followed by a comparative
discussion. Finally, Section~\ref{sec:conclusion} concludes the paper and outlines possible extensions toward broader
intelligent gas-sensing frameworks.

\section{Physical Background and Experimental Basis}
\label{sec:physical_background}
% =========================================================
Before presenting the machine-learning framework, we briefly summarize the experimentally validated mixed-phase SnO-SnO$_2$ CO sensing platform underlying this study~\cite{Gangwar2024}. This section highlights only the physical concepts motivating the regime-wise learning strategy used throughout the paper.

\subsection{Mixed-phase SnO-SnO$_2$ platform and temperature-dependent p-n switching}

The mixed-phase SnO-SnO$_2$ thin-film CO sensor reported in~\cite{Gangwar2024} comprises coexisting p-type SnO and n-type SnO$_2$, forming a heterostructure in which interfacial charge transfer gives rise to temperature-dependent p--n switching and two complementary sensing regimes.
The mixed-phase nature of the sensing platform was verified by XRD and XPS analyses, confirming the coexistence of SnO and SnO$_2$ phases \cite{Mahana2023,Jeong2018}. Hall-effect measurements indicated an overall p-type electronic character of the as-deposited film at room temperature \cite{Granato2013}. Together, these observations support the mixed-phase behavior underlying temperature-dependent p--n switching~\cite{Krishnakumar2008}.

A key result of the experimental study is the observation of temperature-dependent p--n switching. The sensor operates in the p-type regime near 150~$^\circ$C and in the n-type regime near 325~$^\circ$C, with a transition around 225~$^\circ$C. These operating regimes form the basis of the separate machine-learning analyses, enabling direct comparison of concentration inference under distinct sensing behaviors within the same material platform. Because the two regimes exhibit opposite response polarities and different transient characteristics, the p-type and n-type datasets are modeled independently and compared only after separate evaluation.

\subsection{Gas sensing mechanism and resistance-transient interpretation}
The sensing mechanism of metal-oxide-semiconductor gas sensors is governed by oxygen adsorption and target-gas-induced charge transfer. Adsorbed oxygen species withdraw electrons from the oxide surface, forming a space-charge layer that determines the baseline resistance. Consequently, the baseline resistance depends on both the material properties and the operating temperature~\cite{Gangwar2022a}.
Upon CO exposure, the adsorbed oxygen species react with the gas molecules and release electrons back to the oxide surface through the reactions~\cite{Gangwar2022b}:
\begin{align*}
\mathrm{CO}_{(\mathrm{gas})} + \mathrm{O}^{-}_{(\mathrm{ads})} 
&\longrightarrow
\mathrm{CO}_{2(\mathrm{gas})} + e^{-}, \\
2\,\mathrm{CO}_{(\mathrm{gas})} + \mathrm{O}^{2-}_{(\mathrm{ads})}
&\longrightarrow
2\,\mathrm{CO}_{2(\mathrm{gas})} + e^{-}.
\end{align*}
The released electrons modify the carrier concentration and the measured resistance. Because the dominant charge carriers differ between the p-type and n-type regimes, the resistance changes in opposite directions under the same CO exposure.
This polarity reversal is the characteristic feature of the mixed-phase SnO-SnO$_2$ platform. The p-type response dominates at lower operating temperatures, whereas the n-type response dominates at higher temperatures, motivating the separate treatment of the two sensing regimes in this work.

Beyond response polarity, the resistance transient contains concentration-dependent kinetic information from the baseline, gas-exposure, and recovery stages. These stages encode physically meaningful characteristics, including the response amplitude, rise and recovery dynamics, plateau behavior, and integrated response, motivating the use of complete cycle-level transients rather than conventional scalar sensing metrics.

\subsection{Role of the p-n heterojunction and physical justification for regime-wise modeling}
The mixed-phase SnO-SnO$_2$ sensor should not be viewed as a simple combination of independent p-type and n-type materials. Their coexistence forms local p--n heterojunctions, where interfacial charge transfer modifies carrier transport in addition to the surface adsorption process. Consequently, the measured resistance reflects both surface reactions and heterojunction-mediated transport.
This behavior also motivates the learning framework. Combining the p-type and n-type data into a single prediction task would mix opposite response polarities and distinct transient characteristics, reducing physical interpretability. Therefore, the two sensing regimes are modeled independently and compared only after separate analysis.

\subsection{Scope of the present computational extension}

The experimental study of \cite{Gangwar2024} established the material synthesis, structural characterization, and temperature-dependent p--n switching behavior of the mixed-phase SnO-SnO$_2$ sensor. Building upon these measurements, the present work investigates whether cycle-level resistance transients can support:

\begin{enumerate}
\renewcommand{\labelenumi}{(\roman{enumi})}
    \item multi-class concentration classification, i.e., discrimination among discrete CO concentration levels;
    \item continuous concentration regression, i.e., estimation of CO concentration from transient-response data.
\end{enumerate}
To preserve physical interpretability, the proposed feature representation is based primarily on physics-guided transient descriptors, while FFT- and DWT-based features provide complementary information.

% =========================================================
\section{Experimental Data Representation and Dataset Construction}
\label{sec:experimental_data}
% =========================================================

Having established the physical basis of the sensing platform, this section describes how experimentally measured resistance transients are converted into machine-learning-ready samples, introducing the dataset, concentration labels, and leakage-aware grouping strategy.

\subsection{Data source, computational scope, and cycle-wise sample definition}

The machine-learning analysis is based on experimentally measured resistance transients from the mixed-phase SnO-SnO$_2$ CO sensing platform reported in~\cite{Gangwar2024}. The dataset comprises the p-type and n-type operating regimes arising from temperature-dependent p--n switching.
Each experimentally measured sensing cycle is treated as one supervised-learning sample associated with a known CO concentration, preserving sensing-event variability while maintaining a direct connection to the sensing process.
Two regime-specific datasets are constructed from the same mixed-phase SnO-SnO$_2$ platform:
\begin{enumerate}
\renewcommand{\labelenumi}{(\roman{enumi})}
\item p-type regime near $150\,^{\circ}$C;
\item n-type regime near $325\,^{\circ}$C.
\end{enumerate}

Let
\(
\Gamma_i=\{(t_{i,1},R_{i,1}),\ldots,(t_{i,m_i},R_{i,m_i})\}
\)
denote the \(i\)-th experimentally measured resistance transient, where \(t_{i,j}\) and \(R_{i,j}\) denote the sampling time and resistance, respectively. 
Because the number of sampled points $m_i$ may vary across cycles, each transient defines one supervised-learning sample associated with a prescribed CO concentration. A sensing cycle typically comprises:
\begin{enumerate}
\renewcommand{\labelenumi}{(\roman{enumi})}
    \item a pre-exposure baseline segment in air (or an air-dominant state),
    \item an active response segment during CO exposure,
    \item and, when captured within the observation window, a recovery trend after gas removal.
\end{enumerate}

This cycle-wise formulation preserves sensing-event variability while increasing the available training samples.
The supervised-learning tasks are defined at the cycle level:
\begin{align*}
f_{\mathrm{cls}}(\Gamma_i) &\in \mathcal{C}_{\mathrm{disc}} \quad \mbox{ and } \quad
f_{\mathrm{reg}}(\Gamma_i) \in \mathcal{C},
\end{align*}
where \(\mathcal{C}_{\mathrm{disc}}\) denotes the discrete set of concentration classes used for
multi-class classification and \(\mathcal{C}\subset\mathbb{R}_{+}\) denotes the continuous
concentration domain (in ppm) used for regression.

\subsection{Dataset composition and leakage-aware grouped validation}
Each sample \(\Gamma_i\) carries two targets: a categorical concentration label for classification and a numerical concentration value (ppm) for regression. For the p-type sensing regime, the concentration levels are \(500~\mathrm{ppb}, 1~\mathrm{ppm}, 2~\mathrm{ppm}, 5~\mathrm{ppm}, 10~\mathrm{ppm}, 25~\mathrm{ppm}, 50~\mathrm{ppm}\), and \(100~\mathrm{ppm}\). The p-type dataset contains \(192\) cycles distributed across \(8\) concentration classes and \(24\) group-preserved experimental families. The n-type dataset is constructed analogously using the same concentration-label convention and group-preserving design.

Cycle-level samples from the same parent acquisition family, measurement session, or experimental batch are statistically correlated. Splitting such related cycles across training and validation folds may introduce information leakage and produce overly optimistic performance estimates.
To prevent information leakage, each cycle-level sample \(\Gamma_i\) is assigned a group label \(g_i\), with samples sharing the same label belonging to the same experimental family. The dataset is therefore represented as
\(
\mathcal{D}=\left\{(\Gamma_i,y_i,g_i)\right\}_{i=1}^{N},
\)
where \(\Gamma_i\) is the cycle-level resistance transient, \(y_i\) is the concentration label
(discrete or continuous, depending on the task), and \(g_i\) is the group identifier used for
leakage-aware validation.
Group labels are assigned at the parent-acquisition level so that related cycles remain within the same validation fold, preventing correlated samples from being treated as independent observations.
% =========================================================
% =========================================================
\section{Physics-Guided Preprocessing and Feature Construction}\label{sec:physics-guided}
% =========================================================

This section describes how each resistance transient is converted into a structured representation for supervised learning using physics-guided transient descriptors complemented by compact FFT- and DWT-based summaries.

\subsection{Representation pipeline and light preprocessing}

Each sensing cycle $\Gamma_i$ is converted into a structured representation through light preprocessing and descriptor extraction. Four complementary feature configurations are considered:
\begin{itemize}
    \item \textbf{PHYSICS}: physics-guided transient descriptors only,
    \item \textbf{FFT}: compact frequency-domain summaries only,
    \item \textbf{DWT}: compact wavelet-domain summaries only,
    \item \textbf{FUSED}: physics-guided transient descriptors together with FFT and DWT summaries.
\end{itemize}
These  are evaluated individually and in combination to assess the contribution of the transform-domain summaries.

% ---------------------------------------------------------

% ---------------------------------------------------------
Each cycle-level resistance transient is first standardized through light preprocessing. Because experimental cycles may differ in length, noise level, and sampling irregularity, preprocessing improves numerical stability and reproducibility.
Preprocessing consists of temporal ordering, mild smoothing, baseline estimation, common-length resampling, and baseline-relative normalization while preserving the transient morphology.
Let \(x_i\in\mathbb{R}^{L}\) denote the common-length resampled transient associated with the
\(i\)-th cycle, and let \(R_{a,i}\) denote the estimated baseline resistance for that cycle. A
baseline-relative normalized signal is then defined by
\[
\widetilde{x}_{i,\ell}
=
\frac{x_{i,\ell}-R_{a,i}}{\max(|R_{a,i}|,\varepsilon)},
\qquad \ell=1,\dots,L,
\]
where \(\varepsilon>0\) is a small numerical constant introduced only for numerical stability. This normalized signal does not replace the physical resistance values but provides a scale-stable representation for shape analysis and transform summarization.

% ---------------------------------------------------------
\subsection{Physics-guided transient descriptors}
% ---------------------------------------------------------

The primary feature block consists of physics-guided transient descriptors summarizing resistance levels, response measures, temporal characteristics, kinetic slopes, area-based quantities, local statistics, and normalized signal-shape information.
Because the mixed-phase SnO-SnO\(_2\) platform exhibits both p-type and n-type sensing regimes,
the relative-response convention must remain consistent with the operative branch. For the p-type
branch, the relative response feature is computed in the form
\[
SR^{(p)}_{\mathrm{feat}}
=
\frac{|R_g-R_a|}{\max(|R_g|,\varepsilon)}\times 100,
\]
where \(R_a\) denotes the baseline resistance, \(R_g\) the gas-state resistance, and \(\varepsilon>0\) is a small numerical safeguard to avoid division by zero. For the n-type branch, the corresponding relative-response feature is defined analogously as
\[
SR^{(n)}_{\mathrm{feat}}
=
\frac{|R_a-R_g|}{\max(|R_a|,\varepsilon)}\times 100.
\]

Let \(\phi_{\mathrm{phys}}(\Gamma_i)\in\mathbb{R}^{p_{\mathrm{phys}}}\) denotes the
physics-guided descriptor vector extracted from the \(i\)-th sensing cycle, then
\[
\phi_{\mathrm{phys}}(\Gamma_i)
=
\bigl(
\phi_{\mathrm{phys}}^{(1)}(\Gamma_i),\dots,
\phi_{\mathrm{phys}}^{(p_{\mathrm{phys}})}(\Gamma_i)
\bigr)^\top.
\]
In the present implementation, \(p_{\mathrm{phys}}=24\), comprising:
\begin{enumerate}
\renewcommand{\labelenumi}{(\roman{enumi})}
    \item resistance-level descriptors, including \(R_a\), \(R_g\), and absolute response measures;
    \item one regime-consistent relative-response descriptor \(SR_{\mathrm{feat}}\);
    \item temporal descriptors, including response-time, recovery-time, and peak-time summaries;
    \item slope-based kinetic descriptors associated with the rise and recovery phases;
    \item area-based summaries over the localized event window;
    \item local statistical summaries over baseline and plateau (or near-plateau) segments;
    \item normalized-shape descriptors extracted from the cycle-level transient.
\end{enumerate}
The same descriptor structure is used for both sensing regimes, with only the relative-response component following the appropriate convention.

% ---------------------------------------------------------
\subsection{FFT- and DWT-based auxiliary summaries}
% ---------------------------------------------------------

In addition to the physics-guided descriptors, compact FFT- and DWT-based summaries are extracted from the preprocessed transient to capture global frequency and localized multiscale information, respectively.
Let
\(
x_i=(x_i^{(1)},x_i^{(2)},\dots,x_i^{(L)})^\top \in \mathbb{R}^{L}
\)\,
denote the common-length preprocessed transient associated with the \(i\)-th sensing cycle.
For the FFT block, the discrete Fourier transform is written as
\[
\widehat{x}_i^{(k)}
=
\sum_{n=0}^{L-1} x_i^{(n+1)} e^{-2\pi \mathrm{i}kn/L},
\qquad k=0,1,\dots,L-1.
\]
A compact summary of the transform coefficients is retained, yielding the FFT feature map
\[
\phi_{\mathrm{FFT}}(\Gamma_i)\in\mathbb{R}^{p_{\mathrm{FFT}}}.
\]

For the DWT block, a multilevel discrete wavelet decomposition is applied to the same preprocessed
transient, producing approximation and detail coefficients across scales:
\[
x_i
\;\longmapsto\;
\bigl(A_i^{(J)}, D_i^{(J)}, D_i^{(J-1)}, \dots, D_i^{(1)}\bigr).
\]
Similarly, a compact summary of the multiscale coefficients is retained, yielding the DWT feature map
\(
\phi_{\mathrm{DWT}}(\Gamma_i)\in\mathbb{R}^{p_{\mathrm{DWT}}}.
\)

Because gas-sensing transients are nonstationary, multiscale localization captures short-duration changes less naturally represented by a global transform.
Both \(p_{\mathrm{FFT}}\) and \(p_{\mathrm{DWT}}\) are intentionally kept small so that these transform-domain blocks remain auxiliary descriptors. Accordingly, the physics-guided descriptor block forms the core of the framework, with FFT and DWT providing complementary information.
% ---------------------------------------------------------
\subsection{Fused representation}
% ---------------------------------------------------------

To combine complementary information sources, we define the fused feature representation by
concatenating the three descriptor blocks:
\[
\phi_{\mathrm{fused}}(\Gamma_i)
=
\bigl(
\phi_{\mathrm{phys}}(\Gamma_i)^\top,\,
\phi_{\mathrm{FFT}}(\Gamma_i)^\top,\,
\phi_{\mathrm{DWT}}(\Gamma_i)^\top
\bigr)^\top.
\]

Thus, the fused representation augments the physics-guided descriptor block with compact FFT and DWT summaries, enabling their incremental contribution to be assessed directly.
% =========================================================
\section{Supervised Learning Framework, Grouped Validation, and Evaluation Metrics} \label{sec:supervised_framework}
% =========================================================

This section presents the supervised-learning framework for concentration inference, including the learning tasks, model classes, grouped cross-validation strategy, and evaluation metrics used to compare the four feature configurations under leakage-aware validation.

\subsection{Overview of the supervised-learning framework}
% ---------------------------------------------------------

After feature construction, each sensing cycle \(\Gamma_i\) is represented by fixed-dimensional feature vectors. Two supervised-learning tasks are considered:
\begin{enumerate}
\renewcommand{\labelenumi}{(\roman{enumi})}
    \item multi-class concentration classification;
    \item continuous concentration regression.
\end{enumerate}

The same evaluation framework is applied independently to each of the four feature configurations:
\(
\phi_{\mathrm{phys}},\,
\phi_{\mathrm{FFT}},\,
\phi_{\mathrm{DWT}}, \mbox{ and }
\phi_{\mathrm{fused}}.
\)
 In addition, an LSTM baseline is evaluated directly on the raw preprocessed resistance transients to provide a comparison with end-to-end sequence learning.

% ---------------------------------------------------------

Before model fitting, feature vectors are standardized in a fold-wise manner. For
each training fold, the scaling transformation is fitted exclusively on the training subset and
then applied to both the training and validation subsets within that fold. This prevents
distributional leakage and ensures that the validation protocol remains statistically sound.

\subsection{Learning models}
% ---------------------------------------------------------

\paragraph{classification models} For concentration classification, four supervised-learning models are compared:
\begin{enumerate}
\renewcommand{\labelenumi}{(\roman{enumi})}
    \item Random Forest (RF),
    \item Support Vector Machine (SVM) with radial-basis-function kernel,
    \item Multi-Layer Perceptron (MLP).
    \item Long Short-Term Memory (LSTM), trained directly on the raw resistance transients.
\end{enumerate}

These models provide complementary inductive biases for moderate-sized structured feature spaces. RF is a nonlinear ensemble-tree model, SVM is a margin-based kernel classifier, and MLP is a compact feedforward neural network for learning nonlinear feature interactions.

\paragraph{Regression models}
% ---------------------------------------------------------

For continuous concentration estimation, the principal regression model is Random Forest
regression, evaluated in two target-space variants:

\begin{enumerate}
\renewcommand{\labelenumi}{(\roman{enumi})}
    \item linear-target regression;
    \item log-target regression.
\end{enumerate}

In the log-target setting, the model is trained on
\[
\widetilde{y}_i = \log(1+y_i),
\]
with predictions mapped back to the original ppm scale by
\[
\widehat{y}_i = \exp(\widehat{\widetilde{y}}_i)-1.
\]
This transformation reduces the dominance of high-concentration values during training and often improves proportional accuracy across the concentration range.

\subsection{Grouped cross-validation, evaluation metrics, and model comparison}
% ---------------------------------------------------------

A defining feature of the methodology is the use of grouped cross-validation. Because multiple cycle-level samples may originate from the same experimental family, splitting related samples across training and validation sets would introduce information leakage and overestimate predictive performance.
Accordingly, all experiments use grouped K-fold cross-validation. If \(g_i\) denotes the group label of the \(i\)-th sample, all samples sharing that label are kept in the same fold so that no parent experimental family is split across training and validation partitions.

Grouped classification performance is reported using mean Accuracy, Balanced Accuracy, Macro-$F_1$, and Weighted-$F_1$, together with fold-wise standard deviations.
Grouped regression performance is reported using Mean Absolute Error (MAE), Root Mean Squared Error (RMSE), Mean Absolute Percentage Error (MAPE), and the coefficient of determination ($R^2$), averaged across the grouped cross-validation splits.

\paragraph*{Model-comparison logic.}
The benchmarking framework addresses the following scientific questions:

(i) Are the physics-guided transient descriptors alone sufficient for strong concentration inference?

(ii) Do the auxiliary FFT and DWT summaries provide complementary predictive information?

(iii) Does the fused representation consistently improve performance over the individual feature blocks?

Accordingly, all candidate models are evaluated under the same group-preserving protocol to ensure fair, leakage-aware comparison across feature configurations. Fold-wise results are aggregated, and the best-performing models are selected for the analyses in Section~\ref{sec:result}.

\subsection{Workflow summary}
% ---------------------------------------------------------
For reproducibility, Algorithm~\ref{alg:grouped_inference} summarizes the grouped concentration-inference workflow for the feature-engineered pipeline (PHYSICS, FFT, DWT, and FUSED) used for both classification and regression; the LSTM baseline is evaluated separately using the raw preprocessed resistance transients under the same grouped cross-validation protocol.
\begin{algorithm}[!t]
\scriptsize
\caption{Grouped concentration inference workflow}
\label{alg:grouped_inference}
\begin{algorithmic}[1]
\Require Cycle-level dataset
$\mathcal{D}=\{(\Gamma_i,y_i,g_i)\}_{i=1}^N$, where $\Gamma_i$ denotes the $i$-th resistance transient,
$y_i$ the corresponding concentration target (class label or ppm value), and $g_i$ the group label

\Ensure Grouped performance summaries across PHYSICS, FFT, DWT, and FUSED feature representations

\State Construct, for each cycle $\Gamma_i$, the four feature representations
\[
\phi_{\mathrm{phys}}(\Gamma_i),\quad
\phi_{\mathrm{FFT}}(\Gamma_i),\quad
\phi_{\mathrm{DWT}}(\Gamma_i),\quad
\phi_{\mathrm{fused}}(\Gamma_i).
\]

\For{each feature configuration $\phi \in \{\phi_{\mathrm{phys}},\phi_{\mathrm{FFT}},\phi_{\mathrm{DWT}},\phi_{\mathrm{fused}}\}$}

    \State Form the feature matrix $X_{\phi}$

    \State Train the selected learning model on the training subset.

    \For{each candidate learning model under comparison}

        \State Partition the dataset using grouped $K$-fold cross-validation with group labels $\{g_i\}_{i=1}^N$

        \For{each grouped fold}

            \State Split the data into training and validation subsets at the group level

            \State Fit the standardization transform using only the training subset

            \State Apply the fitted transform to both training and validation subsets

            \If{log-target regression is used}
                \State Transform training targets as
                $\widetilde{y}_i=\log(1+y_i)$
            \EndIf

            \State Train the selected learning model on the training subset.

            \State Predict concentration targets on the validation subset

            \If{log-target regression is used}
                \State Map predictions back to the original scale via
                $\widehat{y}_i=\exp(\widehat{\widetilde{y}}_i)-1$
            \EndIf

            \If{classification}
                \State Compute Accuracy, Balanced Accuracy, Macro-$F_1$ and Weighted-$F_1$
            \Else
                \State Compute MAE, RMSE, MAPE, and $R^2$
            \EndIf

        \EndFor

        \State Aggregate the fold-wise metrics for the current learning model and feature configuration

    \EndFor

\EndFor

\State Compare grouped results across feature configurations and learning models

\State Record the strongest grouped model for each sensing regime and learning task

\end{algorithmic}
\end{algorithm}

% =========================================================
% =========================================================
\section{Results and Discussion}
\label{sec:result}
% =========================================================

This section presents the machine-learning results for the p-type and n-type sensing regimes under the leakage-aware grouped-validation protocol. Results are organized around representative transient behavior, comparative feature-set performance, best-model diagnostics, and feature-importance analysis for the four feature configurations. Classification performance is evaluated using Accuracy, Balanced Accuracy, Macro-\(F_1\), and Weighted-\(F_1\), whereas regression performance is evaluated using MAE, RMSE, MAPE, and \(R^2\).
% ---------------------------------------------------------
\subsection{Representative transient-response behavior}
\label{subsec:representative_transients}
% ---------------------------------------------------------
\begin{figure}[!t]
\centering
\begin{subfigure}[t]{0.49\columnwidth}
    \centering
    \includegraphics[width=\linewidth]{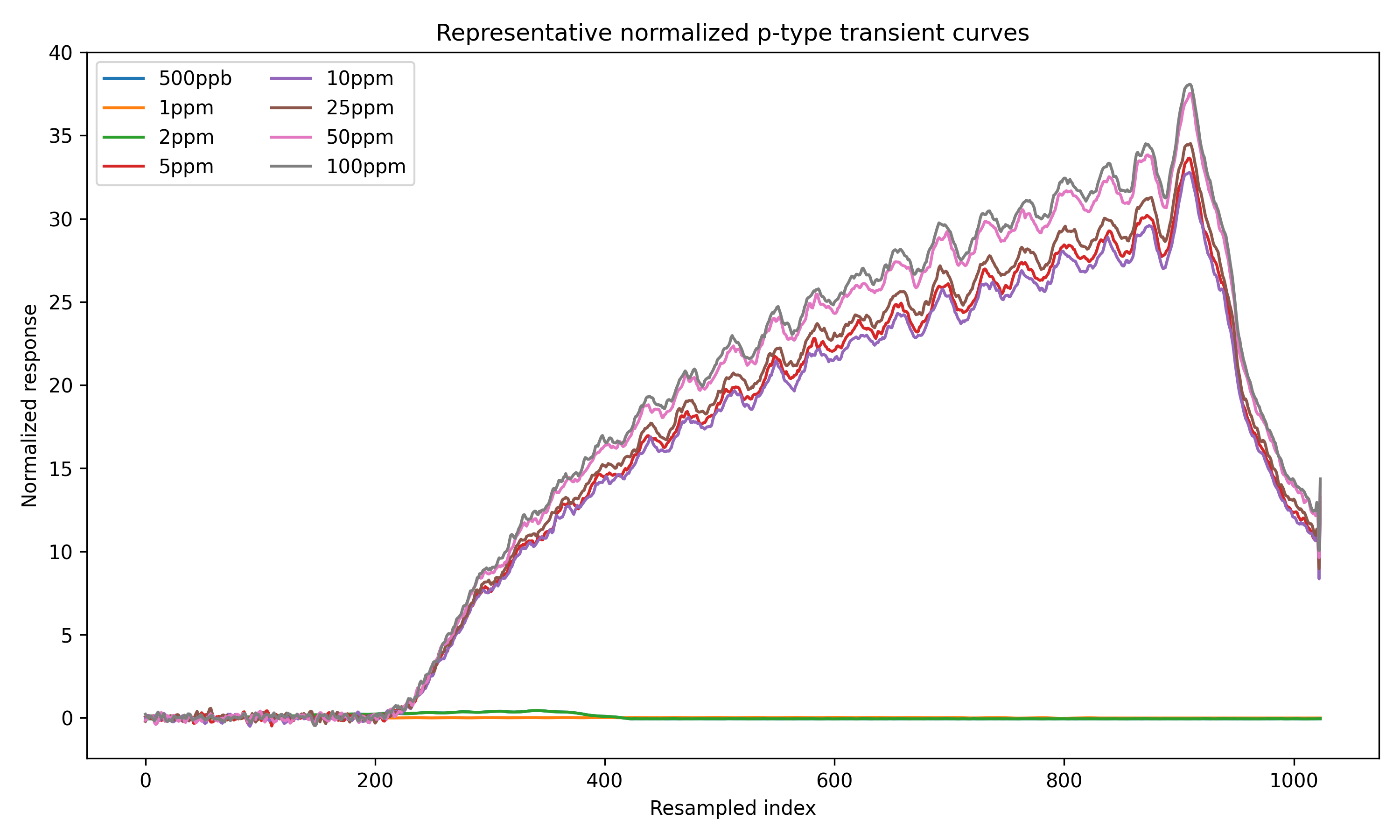}
    \caption{p-type regime.}
    \label{fig:representative_transients_p}
\end{subfigure}
\hfill
\begin{subfigure}[t]{0.49\columnwidth}
    \centering
    \includegraphics[width=\linewidth]{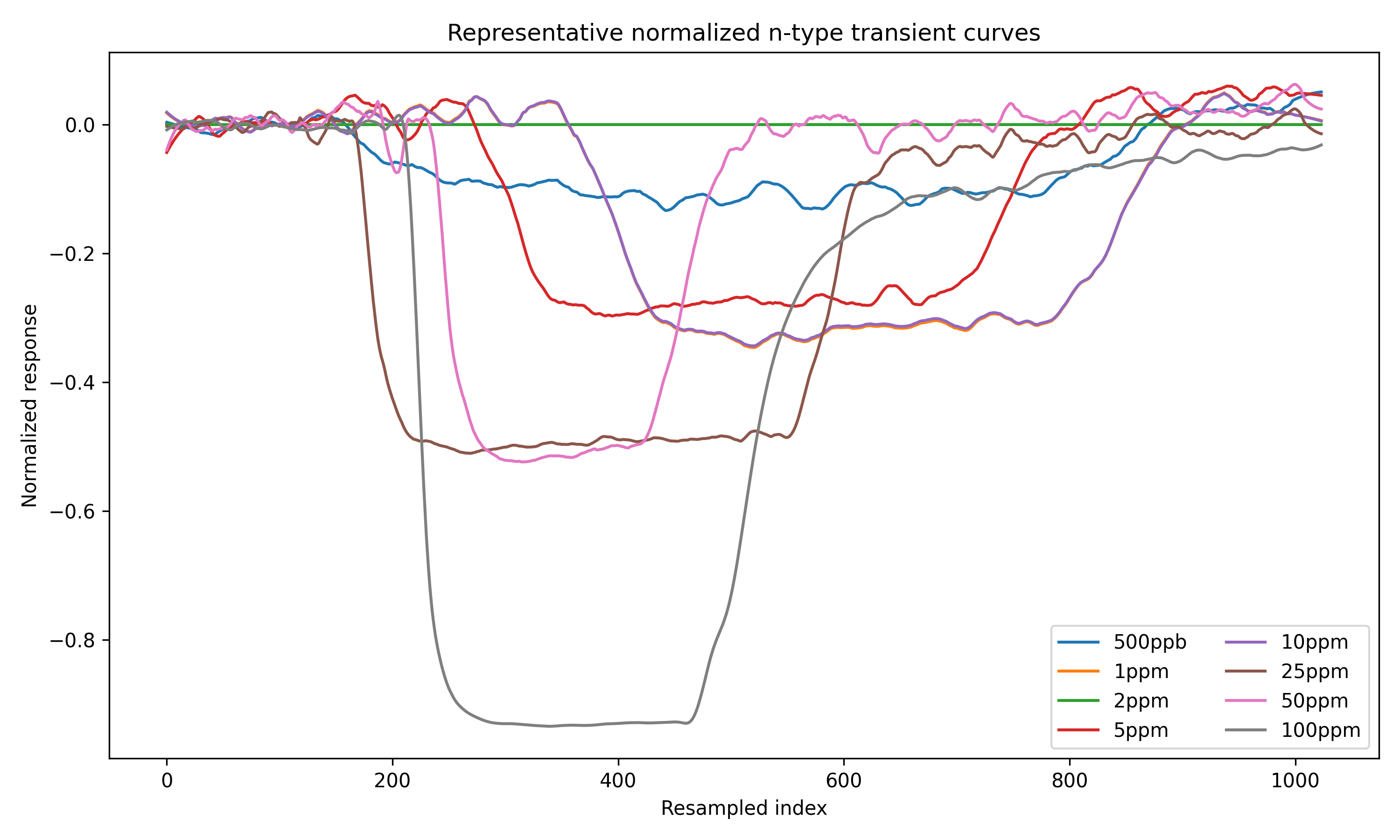}
    \caption{n-type regime.}
    \label{fig:representative_transients_n}
\end{subfigure}
\caption{Representative cycle-level resistance transients.}
\label{fig:representative_transients}
\end{figure}
Representative cycle-level resistance transients for the p-type and n-type sensing regimes are shown in Fig.~\ref{fig:representative_transients}. They confirm that concentration information is encoded not only in the overall response magnitude but also in the full temporal morphology of the sensing cycle, including onset behavior, local slope structure, plateau characteristics, and recovery dynamics. This motivates the use of physics-guided transient descriptors while explaining why compact transform-domain summaries provide complementary information in the fused representation.

% ---------------------------------------------------------
\subsection{Grouped classification performance}
\label{subsec:grouped_classification}
% ---------------------------------------------------------

We first evaluate discrete concentration classification under the leakage-aware grouped-validation protocol. The objective is to identify the strongest classifier in each sensing regime and compare the four feature configurations.

% ---------------------------------------------------------
\subsubsection{Main grouped classification results}
% ---------------------------------------------------------
Table~\ref{tab:classification_main} summarizes the best grouped classification results for the two sensing regimes. The p-type branch achieves the highest classification performance, whereas the n-type branch remains predictive but with a lower classification ceiling.

\begin{table}[!t]
\centering
\caption{Main grouped classification results for the p-type and n-type sensing regimes.}
\label{tab:classification_main}
\scriptsize
\setlength{\tabcolsep}{4pt}
\begin{tabular}{llccc}
\toprule
Regime & Best model & Accuracy & Balanced Accuracy & Macro-\(F_1\) \\
\midrule
p-type & FUSED + RF  & 0.9650 & 0.9650 & 0.9652 \\
n-type & FUSED + MLP & 0.7850 & 0.7850 & 0.7716 \\
\bottomrule
\end{tabular}
\end{table}
The p-type branch attains the strongest grouped classification result of the study:
\(
\mathrm{Accuracy}=96.50\%\),\,
\(\mathrm{Balanced\ Accuracy}=96.50\%\),\,
\(\mathrm{Macro}\text{-}F_1=96.52\%.
\)
This indicates that the p-type transient responses form sharply separated concentration-dependent clusters, making the p-type regime especially well suited to concentration-level classification.
By contrast, the best n-type classifier is the fused MLP model, with
\(
\mathrm{Accuracy}=\!78.5\%\),\,
\(\mathrm{Balanced\ Accuracy}=78.5\%\),\,
\(\mathrm{Macro}\text{-}F_1=77.16\%.
\)
Although lower than the p-type result, it remains a strong grouped-validation outcome, indicating weaker concentration-level separation.

% ---------------------------------------------------------
\subsubsection{Comparative behavior across PHYSICS, FFT, DWT, and FUSED representations}
% ---------------------------------------------------------
\begin{figure}[!t]
\centering
\begin{subfigure}[t]{0.49\columnwidth}
    \centering
    \includegraphics[width=\linewidth]{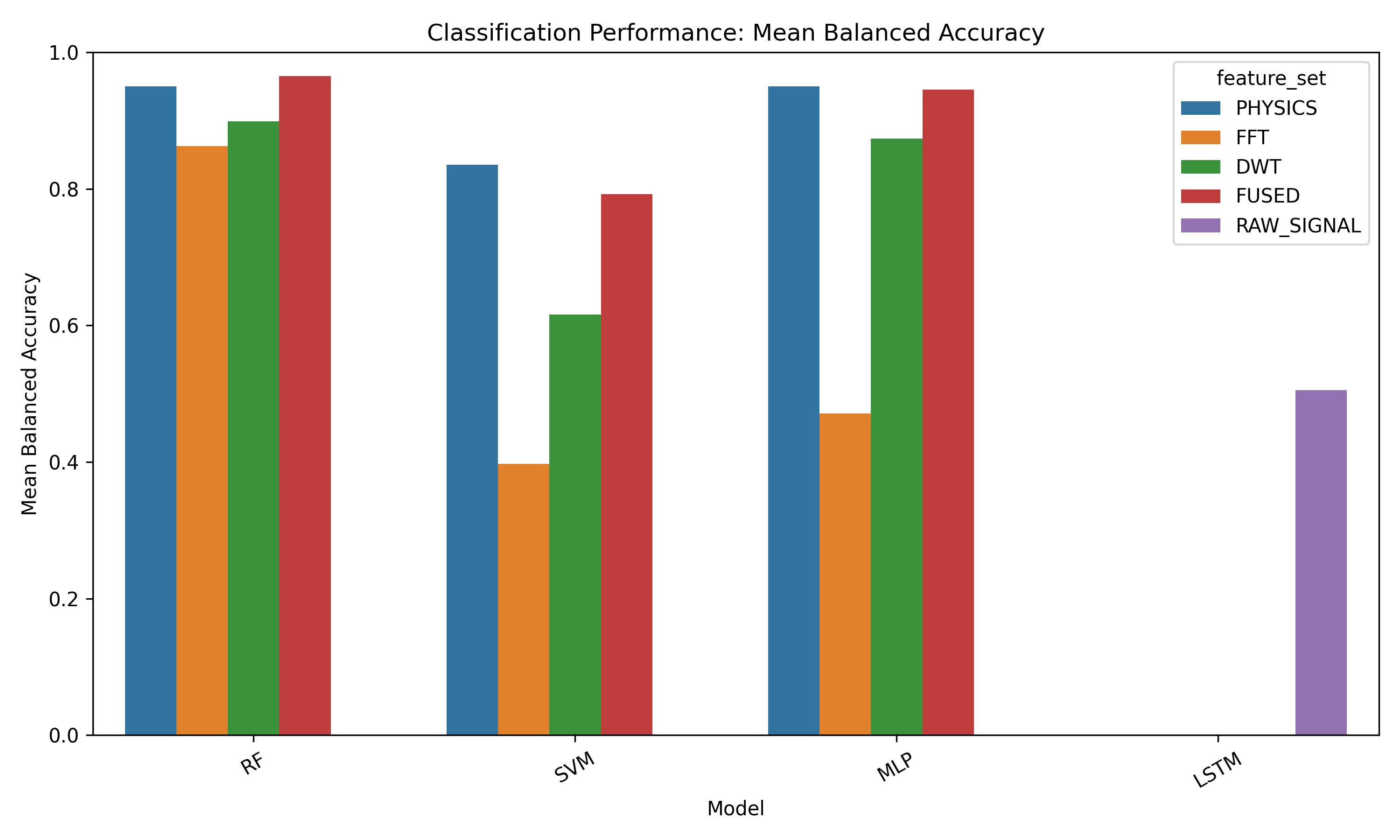}
    \caption{p-type regime.}
    \label{fig:classification_comparison_p}
\end{subfigure}
\hfill
\begin{subfigure}[t]{0.49\columnwidth}
    \centering
    \includegraphics[width=\linewidth]{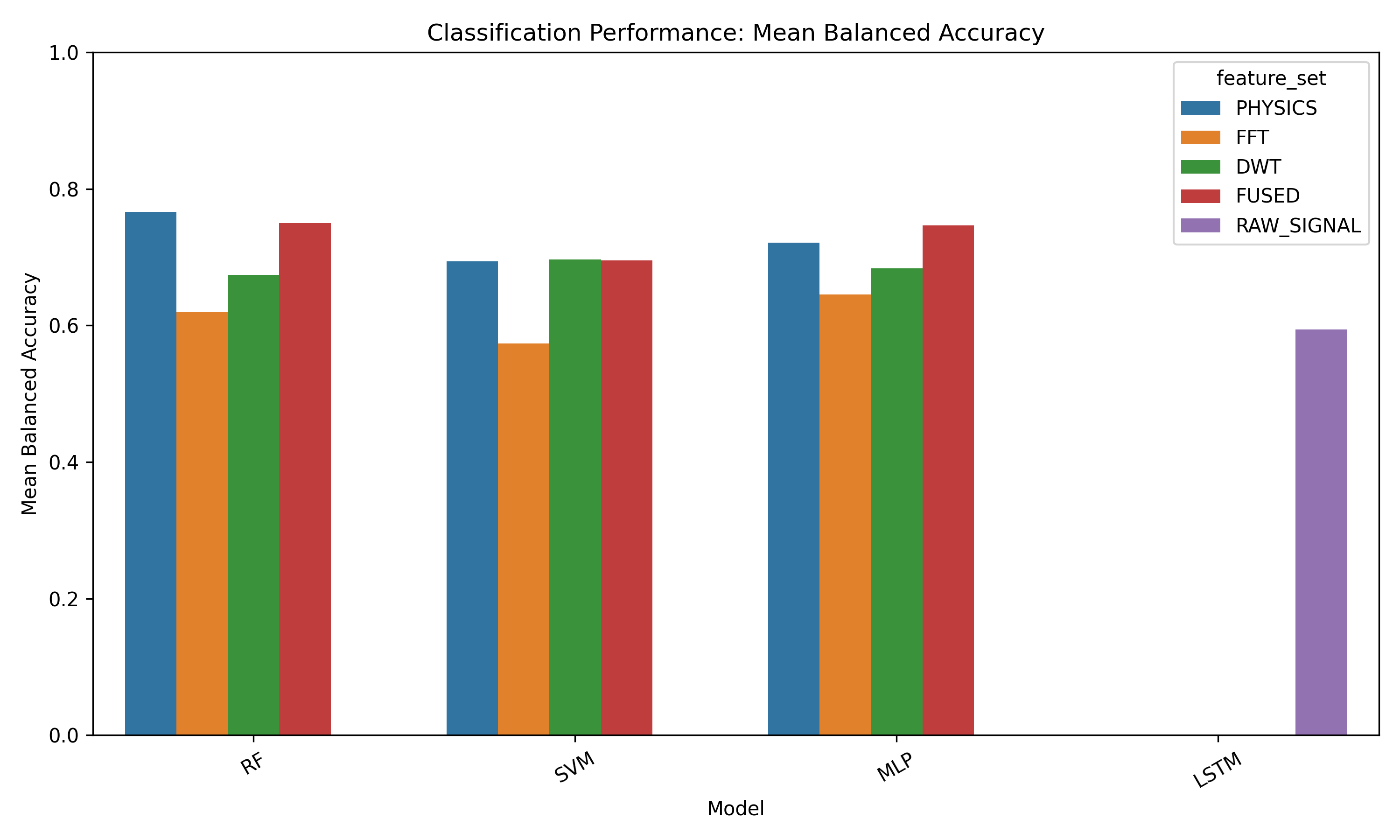}
    \caption{n-type regime.}
    \label{fig:classification_comparison_n}
\end{subfigure}
\caption{Grouped classification performance comparison.}
\label{fig:classification_comparison_side}
\end{figure}

\begin{figure}[!t]
\centering

\begin{subfigure}[t]{0.49\columnwidth}
    \centering
    \includegraphics[width=\linewidth]{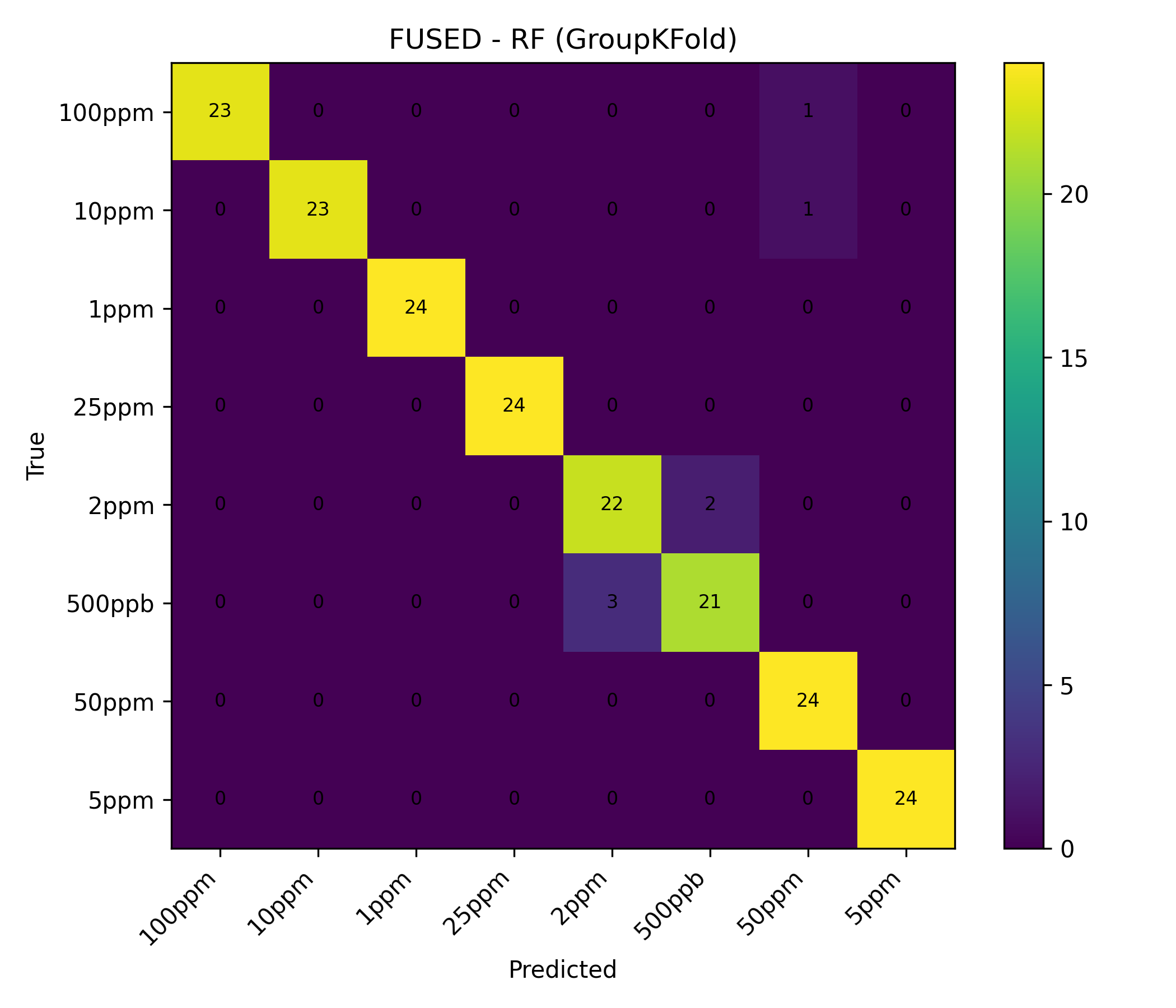}
   \caption{Best grouped p-type classifier.}
    \label{fig:classification_confusion_p}
\end{subfigure}
\hfill
\begin{subfigure}[t]{0.49\columnwidth}
    \centering
    \includegraphics[width=\linewidth]{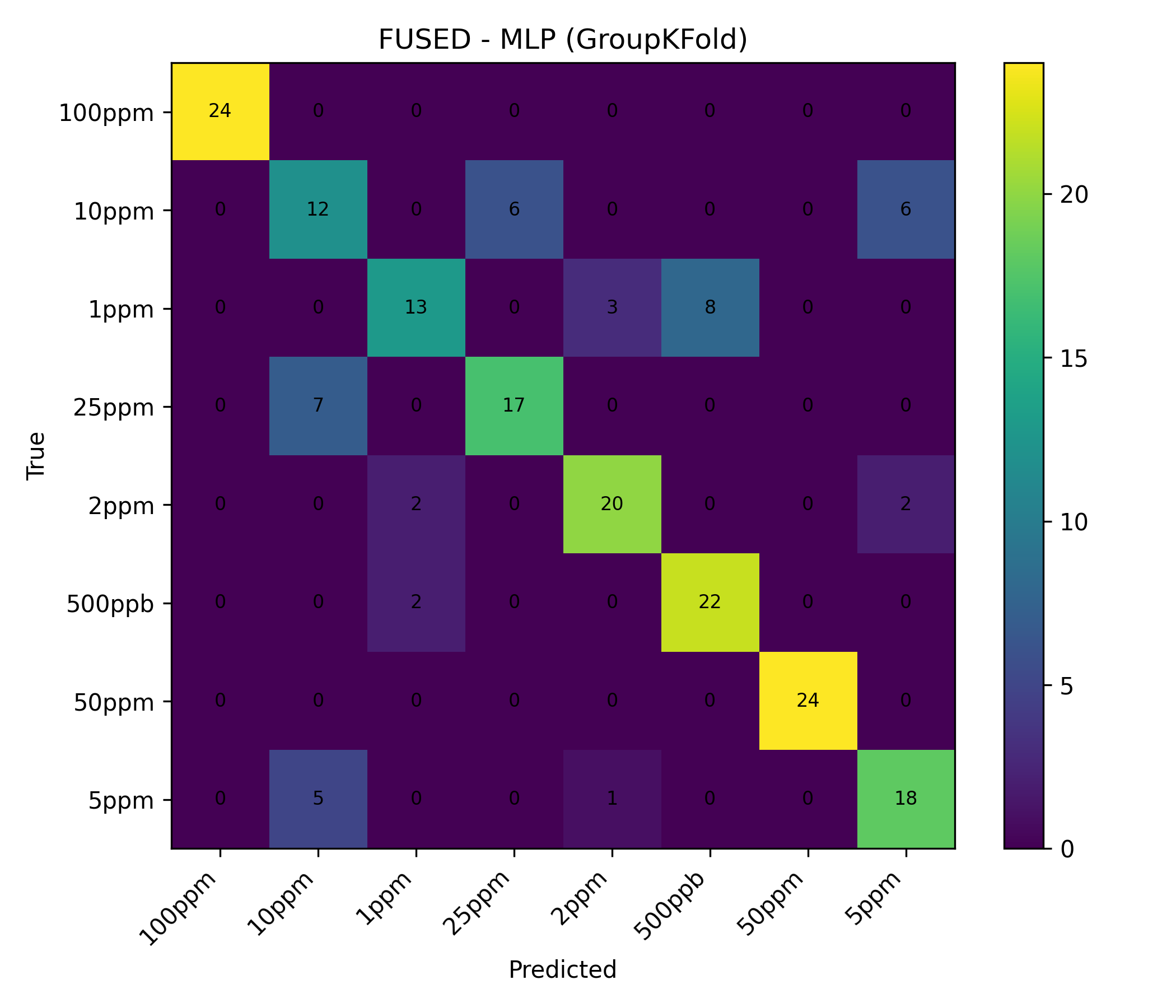}
    \caption{Best grouped n-type classifier.}
    \label{fig:classification_confusion_n}
\end{subfigure}

\caption{Confusion matrices of the best grouped classifiers.}
\label{fig:classification_confusions_side}
\end{figure}
Fig.~\ref{fig:classification_comparison_side} shows that the PHYSICS representation is already highly informative, whereas the transform-only representations are weaker and the FUSED representation consistently performs best. The DWT block is generally more competitive than FFT, consistent with the nonstationary nature of gas-sensing transients. The LSTM baseline trained directly on the raw resistance transients achieves lower classification performance than the best feature-based models, indicating that the proposed engineered feature representations provide more discriminative information for this dataset.
Fig.~\ref{fig:classification_confusions_side} provides a detailed view of the classification errors. In the p-type branch, the dominant diagonal structure indicates that most errors are limited
to nearby concentration levels, consistent with a well-ordered concentration-response hierarchy. In
the n-type branch, the broader off-diagonal spread indicates greater overlap between neighboring
classes, which explains the lower grouped classification ceiling.

\subsection{Machine-learning model configuration}
Table~\ref{tab:hyperparameters} summarizes the fixed hyperparameter settings used throughout all experiments.
\begin{table}[!t]
\caption{Hyperparameter settings used in this study.}
\label{tab:hyperparameters}
\centering
\scriptsize
\setlength{\tabcolsep}{3pt}
\begin{tabular}{lll}
\toprule
\textbf{Model} & \textbf{Hyperparameter} & \textbf{Value} \\
\midrule

RF (Classification)
& Number of trees & 500 \\
& Maximum tree depth & None \\
& Minimum samples per leaf & 1 \\
& Random seed & 42 \\

\midrule

SVM
& Kernel function & RBF \\
& Regularization parameter ($C$) & 10 \\
& Kernel coefficient ($\gamma$) & scale \\
& Feature scaling & StandardScaler \\

\midrule

MLP
& Hidden layer sizes & (256, 128, 64) \\
& Activation function & ReLU \\
& Optimization algorithm & Adam \\
& L2 regularization ($\alpha$) & $10^{-4}$ \\
& Maximum training iterations & 4000 \\
& Random seed & 42 \\

\midrule

LSTM
& Input representation & Raw resistance transient \\
& LSTM hidden units & (32, 16) \\
& Dense layer & 64 \\
& Dropout rate & 0.30 \\
& Optimizer & Adam \\
& Learning rate & $10^{-3}$ \\
& Batch size & 16 \\
& Maximum training epochs & 100 \\
& Early stopping & Enabled \\

\midrule

RF (Regression)
& Number of trees & 600 \\
& Maximum tree depth & None \\
& Random seed & 42 \\

\bottomrule
\end{tabular}
\end{table}
% ---------------------------------------------------------
\subsection{Grouped regression performance}
\label{subsec:grouped_regression}
% ---------------------------------------------------------

We next evaluate continuous concentration regression under the same leakage-aware grouped-validation framework. The objective is to identify the strongest regressor in each sensing regime and compare the four feature configurations using both linear- and log-target formulations.

% ---------------------------------------------------------
\subsubsection{Main grouped regression results}
% ---------------------------------------------------------

\begin{table}[!t]
\caption{Main grouped regression results for the p-type and n-type sensing regimes.}
\label{tab:regression_main}
\centering
\scriptsize
\setlength{\tabcolsep}{3pt}
\begin{tabular}{llcccc}
\toprule
Regime & Best model & MAE & RMSE & MAPE (\%) & $R^2$ \\
\midrule
p-type & FUSED + RFReg. (linear) & 1.4924 & 4.3434 & 27.5964 & 0.9824 \\
n-type & FUSED + RFReg. (log)    & 1.4757 & 2.9697 & 24.2273 & 0.9918 \\
\bottomrule
\end{tabular}
\end{table}
Table~\ref{tab:regression_main} summarizes the best grouped regression results for the two sensing regimes. Unlike classification, the n-type branch achieves the strongest overall regression performance.
For the p-type branch, the fused Random Forest regressor with the linear target achieves
\(
\mathrm{MAE}=1.492\ \mathrm{ppm}\),\,
\(\mathrm{RMSE}=4.343\ \mathrm{ppm}\),\,
\(R^2=0.9824.
\)
This strong grouped-validation result shows that the p-type branch is effective for both concentration classification and continuous regression.
For the n-type branch, the fused Random Forest regressor with the log-target achieves
\(
\mathrm{MAE}=1.476\ \mathrm{ppm}\),\,
\(\mathrm{RMSE}=2.970\ \mathrm{ppm}\),\,
\(\mathrm{MAPE}=24.23\%\),\,
\(R^2=0.9918.
\)
This grouped-validation result surpasses the best p-type regression performance, revealing a clear contrast with the classification results.

% ---------------------------------------------------------
\subsubsection{Comparative behavior across PHYSICS, FFT, DWT, and FUSED representations}
% ---------------------------------------------------------

\begin{figure}[!t]
\centering

\begin{subfigure}[t]{0.49\columnwidth}
    \centering
    \includegraphics[width=\linewidth]{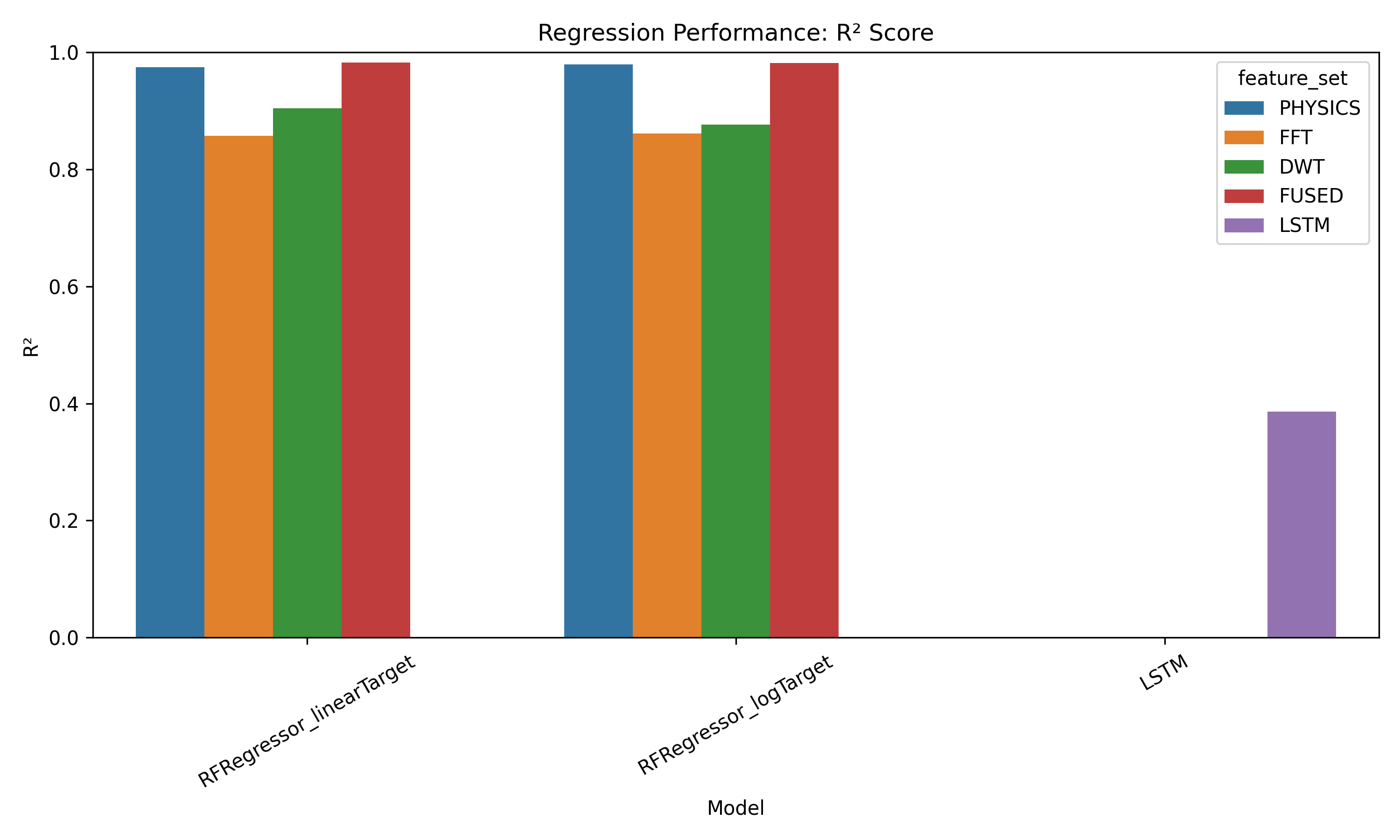}
    \caption{p-type regime.}
    \label{fig:regression_comparison_p}
\end{subfigure}
\hfill
\begin{subfigure}[t]{0.49\columnwidth}
    \centering
    \includegraphics[width=\linewidth]{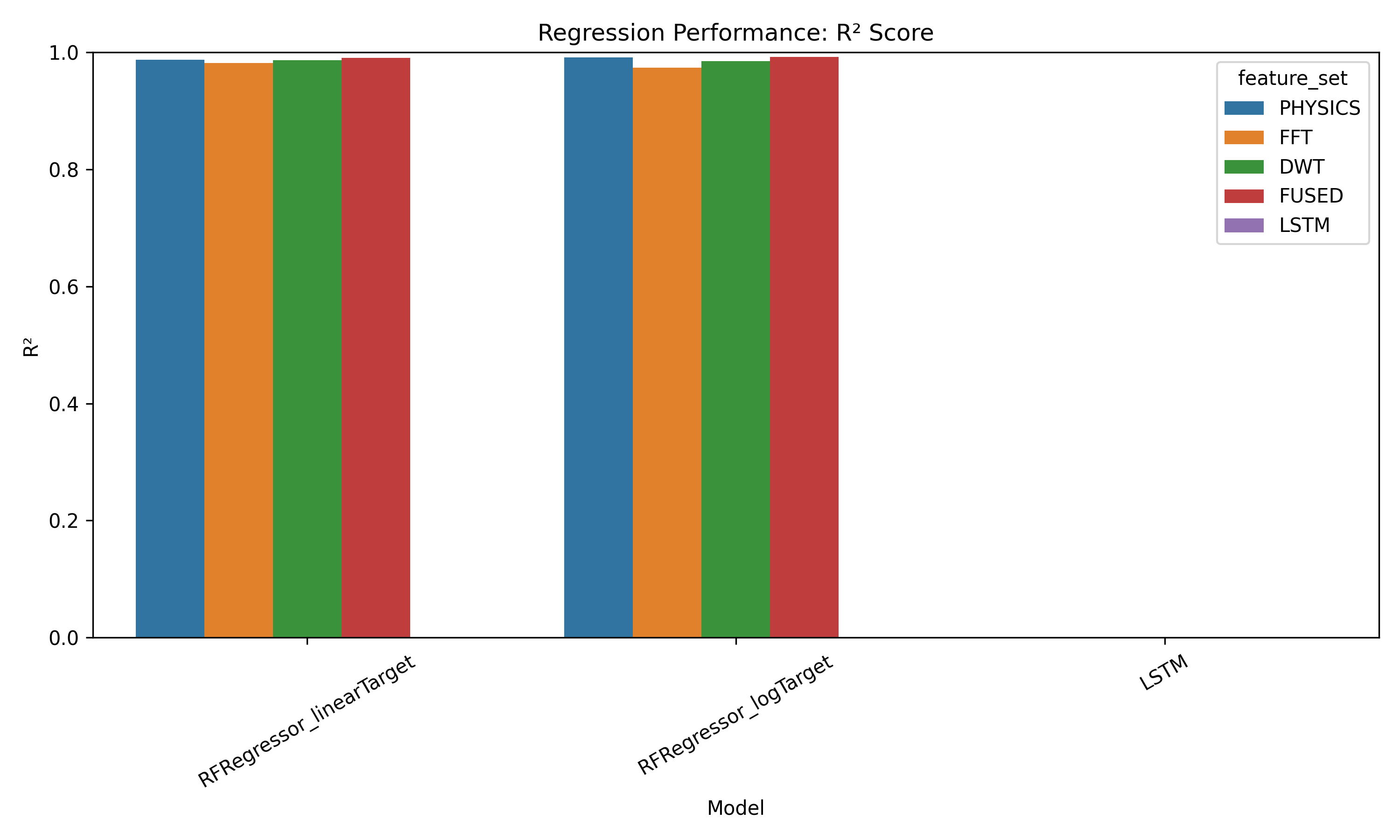}
    \caption{n-type regime.}
    \label{fig:regression_comparison_n}
\end{subfigure}

\caption{Grouped regression performance comparison.}
\label{fig:regression_comparison_side}
\end{figure}
Fig.~\ref{fig:regression_comparison_side} compares the grouped regression performance obtained using the PHYSICS, FFT, DWT, FUSED, and LSTM representations for both sensing regimes. Consistent with the classification results, the PHYSICS representation already captures most of the predictive structure, whereas the FUSED representation consistently achieves the strongest overall performance. The DWT representation generally outperforms the FFT representation, indicating that multiscale transient characteristics provide more informative concentration cues than global frequency summaries. In contrast, the LSTM baseline trained directly on the raw resistance transients performs noticeably worse than the feature-engineered approaches, particularly for the n-type sensing regime. These results demonstrate that the proposed physics-guided feature representations extract more informative concentration-related features from the available experimental data than direct end-to-end sequence learning.
Model fidelity is further assessed using prediction-agreement and concentration-dependent error diagnostics.
\begin{figure}[!t]
\centering

\begin{subfigure}[t]{0.49\columnwidth}
    \centering
    \includegraphics[width=\linewidth]{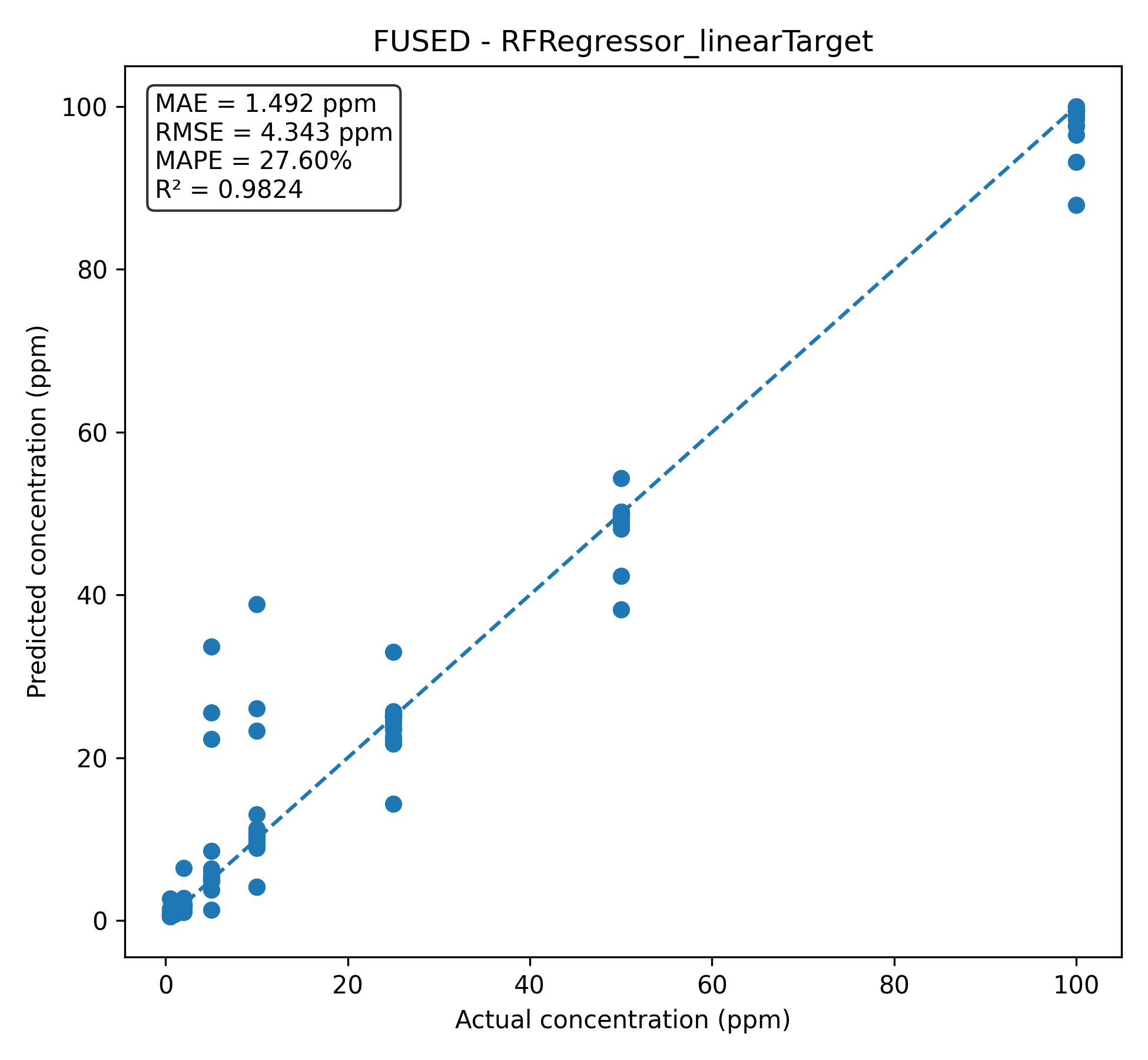}
    \caption{Best p-type grouped regressor.}
    \label{fig:regression_pred_actual_p}
\end{subfigure}
\hfill
\begin{subfigure}[t]{0.49\columnwidth}
    \centering
    \includegraphics[width=\linewidth]{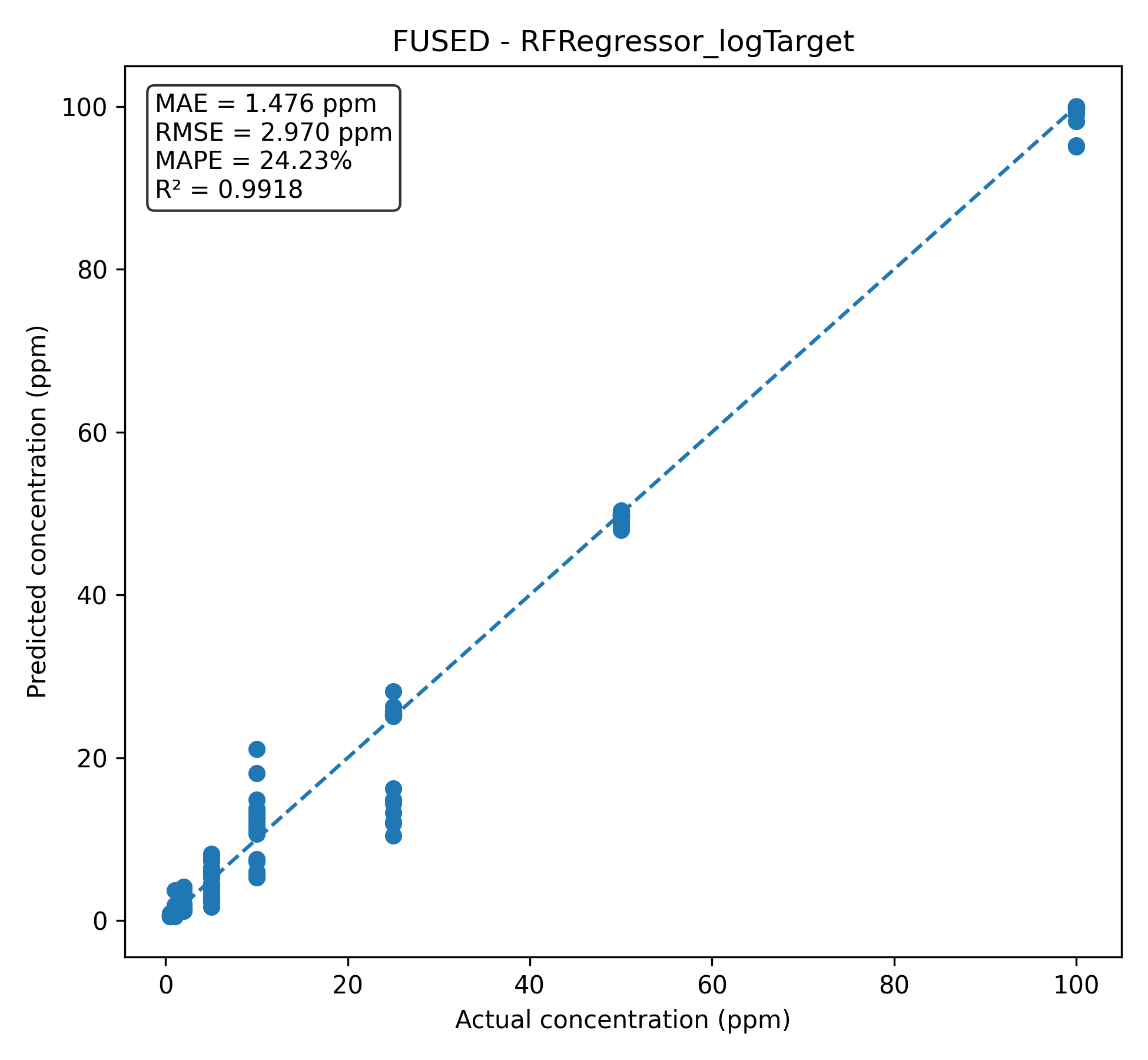}
    \caption{Best n-type grouped regressor.}
    \label{fig:regression_pred_actual_n}
\end{subfigure}

\caption{Predicted-versus-actual concentration plots for the best grouped regressors.}
\label{fig:regression_pred_vs_actual_side}
\end{figure}

Fig.~\ref{fig:regression_pred_vs_actual_side} provides a direct assessment of regression fidelity through predicted-versus-actual concentration plots. In both sensing regimes, the predicted concentrations closely follow the identity line, demonstrating excellent agreement with the experimentally measured concentrations under leakage-aware grouped validation. The p-type and n-type branches both achieve accurate concentration estimation; however, the n-type branch exhibits a visibly tighter alignment with the identity line, consistent with its lower MAE and RMSE values and \(R^2\). These qualitative observations are consistent with the quantitative results reported in Table~\ref{tab:regression_main}, confirming that the proposed physics-guided feature representations enable accurate and robust continuous concentration estimation across both sensing regimes.

\begin{figure}[!t]
\centering

\begin{subfigure}[t]{0.49\columnwidth}
    \centering
    \includegraphics[width=\linewidth]{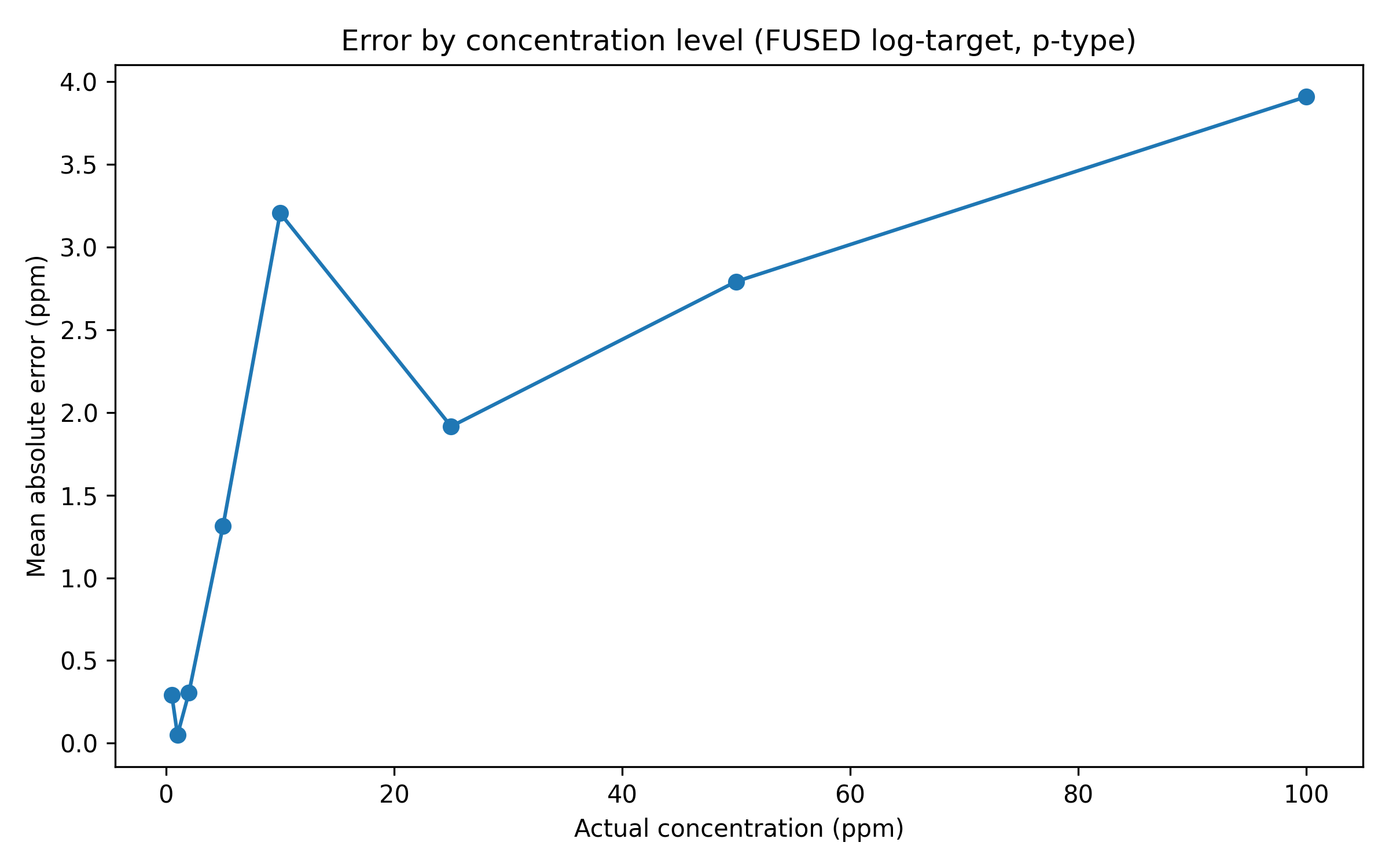}
    \caption{p-type regime.}
    \label{fig:regression_error_conc_p}
\end{subfigure}
\hfill
\begin{subfigure}[t]{0.49\columnwidth}
    \centering
    \includegraphics[width=\linewidth]{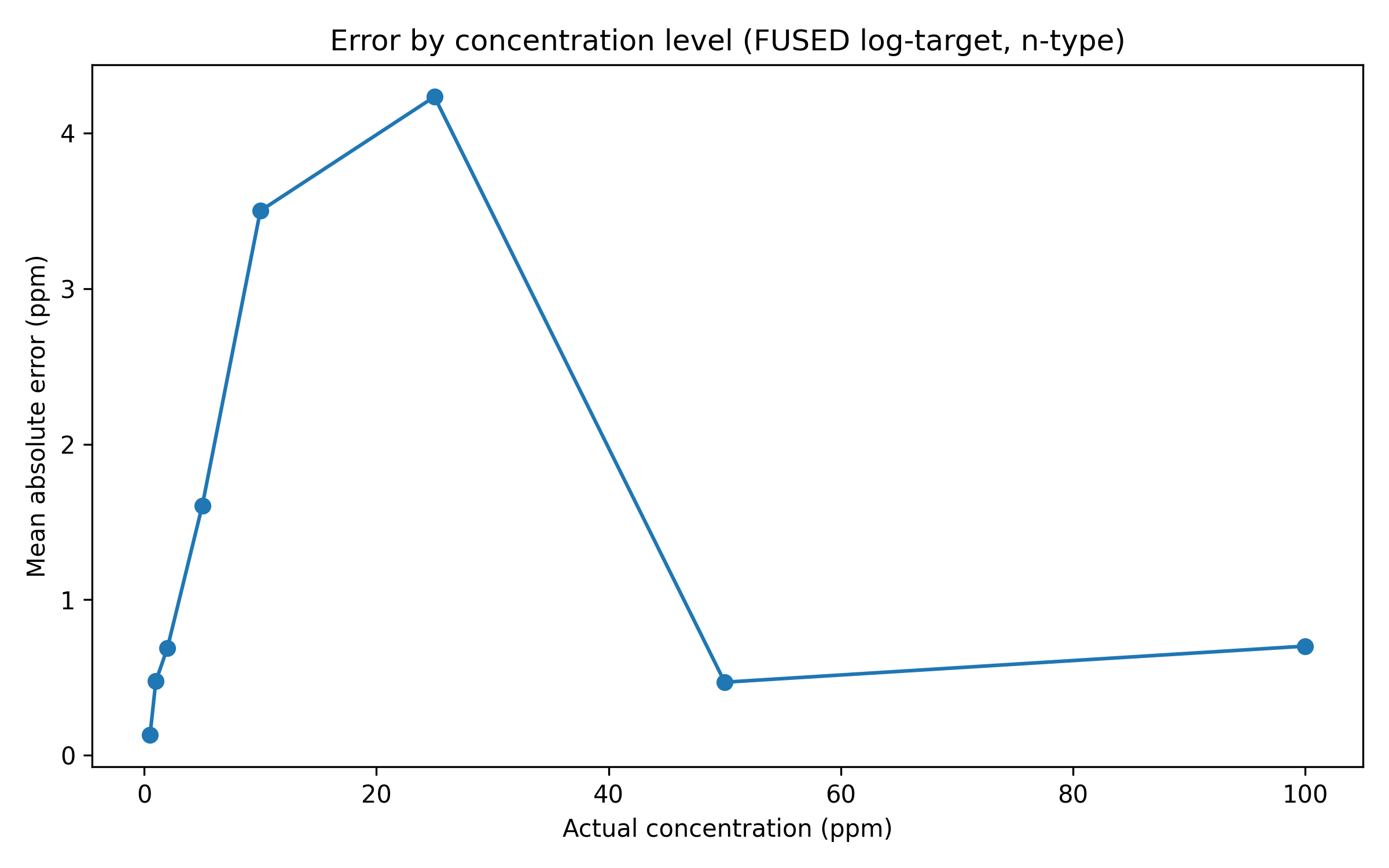}
    \caption{n-type regime.}
    \label{fig:regression_error_conc_n}
\end{subfigure}

\caption{Concentration-wise grouped regression error for the selected FUSED regressors.}
\label{fig:regression_error_by_concentration_side}
\end{figure}
Fig.~\ref{fig:regression_error_by_concentration_side} provides a complementary view of concentration-dependent regression difficulty by showing the mean absolute error at each concentration level. In both sensing regimes, the error varies across the studied concentration range, indicating that some concentration levels are more difficult to estimate than others under grouped validation. This behavior is consistent with concentration-dependent variation in signal amplitude, transient-shape overlap, and baseline sensitivity. These diagnostics identify concentration ranges with relatively larger prediction errors.

% ---------------------------------------------------------
\subsection{Feature correlation, feature importance, and physical interpretability}
\label{subsec:feature_importance}
\begin{figure}[!t]
\centering

\begin{subfigure}[t]{0.49\columnwidth}
    \centering
    \includegraphics[width=\linewidth]{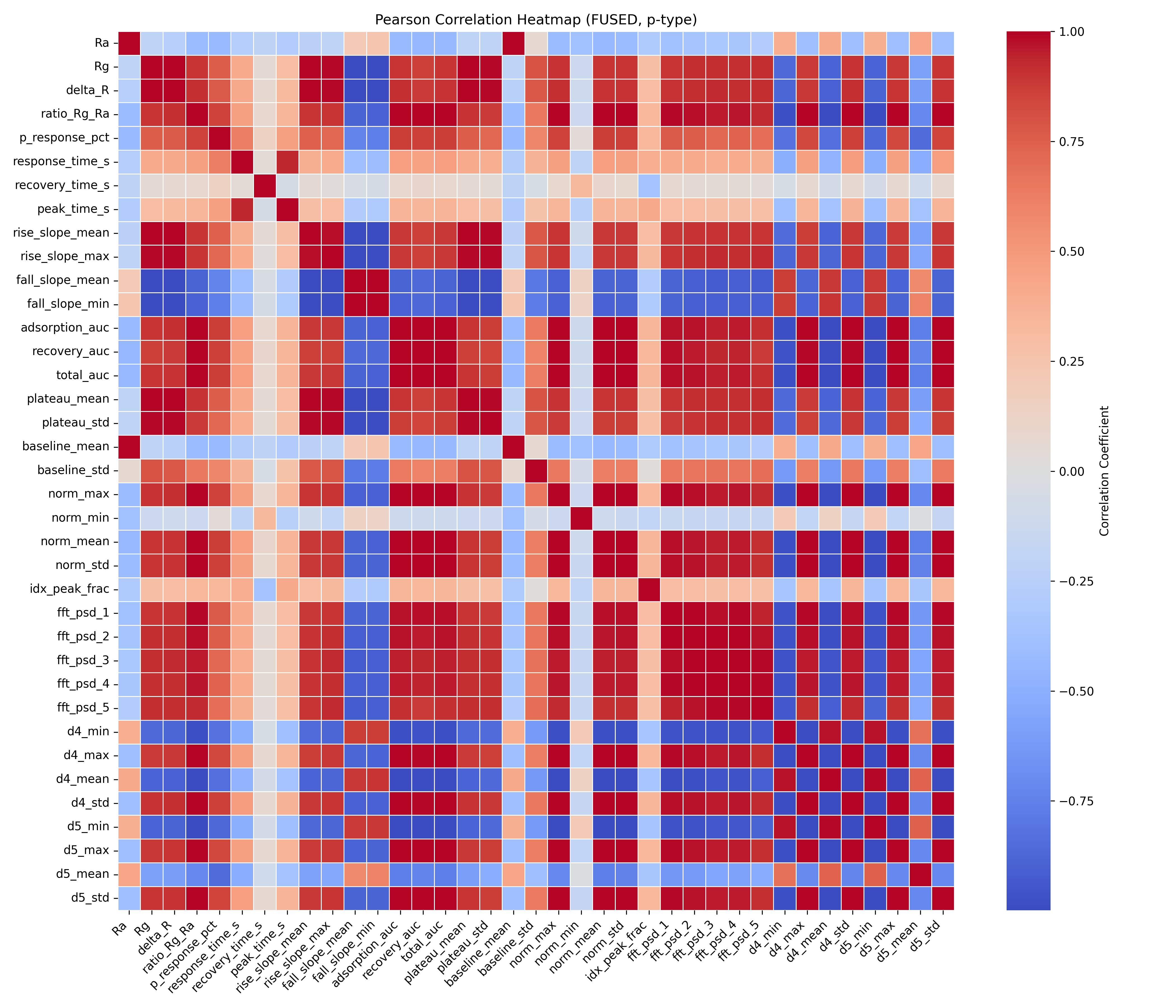}
    \caption{p-type regime.}
\end{subfigure}
\hfill
\begin{subfigure}[t]{0.49\columnwidth}
    \centering
    \includegraphics[width=\linewidth]{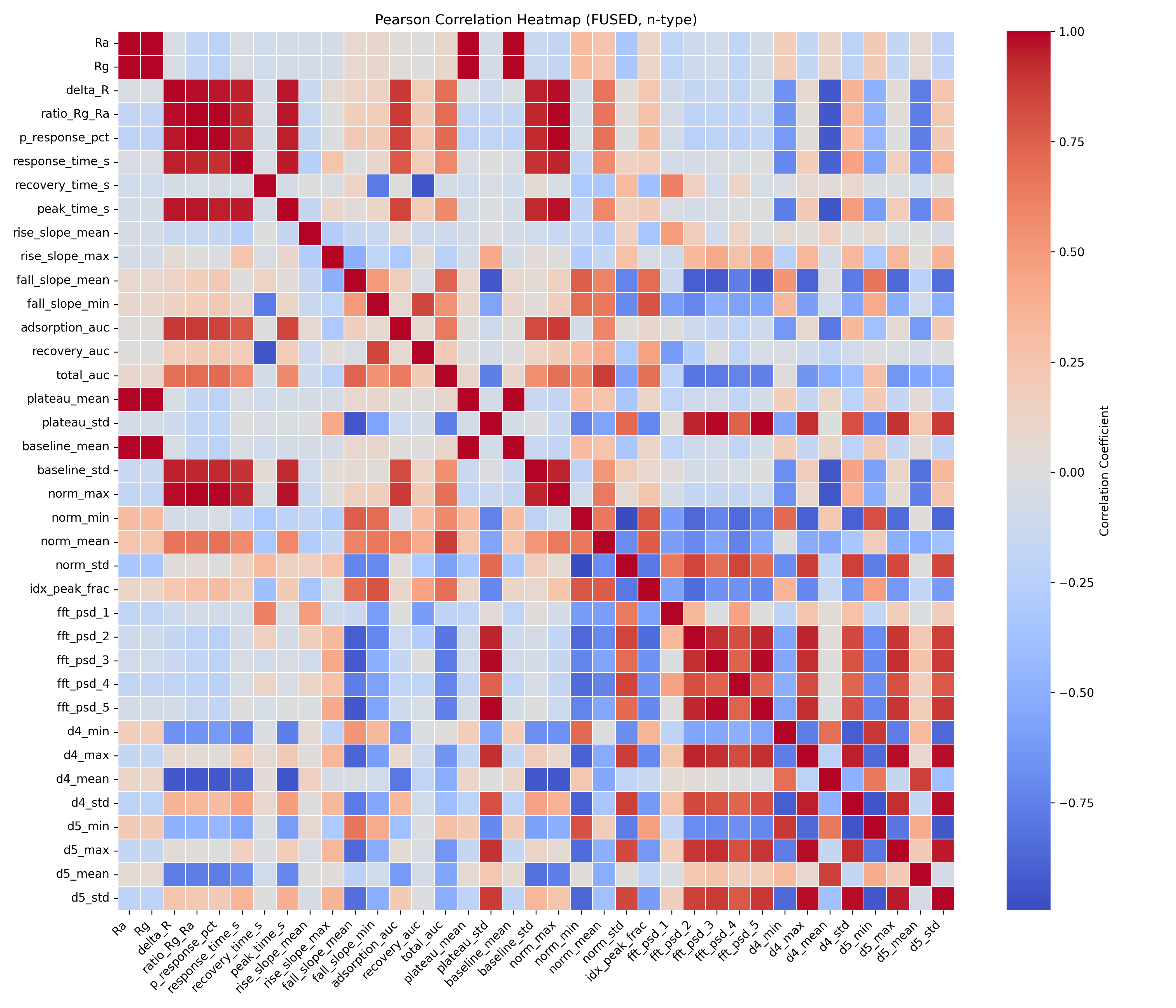}
    \caption{n-type regime.}
\end{subfigure}

\caption{Pearson correlation heatmaps of the FUSED feature.}
\label{fig:feature_correlation}
\end{figure}

% ---------------------------------------------------------
Before examining feature importance, we investigate the statistical dependence among the engineered features using the Pearson correlation coefficient. Fig.~\ref{fig:feature_correlation} presents the Pearson correlation matrices of the FUSED feature representation for the p-type and n-type sensing regimes.

Because the strongest regression models in both sensing regimes are Random Forest based, feature-importance analysis provides a direct connection between predictive performance and physical interpretability by identifying the transient descriptors most relevant to concentration estimation. To assess the robustness of these findings, we compare the impurity-based Random Forest feature importance with Permutation Feature Importance (PFI), which measures feature relevance through the decrease in predictive performance after randomly permuting each feature.

\begin{figure*}[!t]
\centering

\begin{subfigure}[t]{0.245\textwidth}
    \centering
    \includegraphics[width=\linewidth]{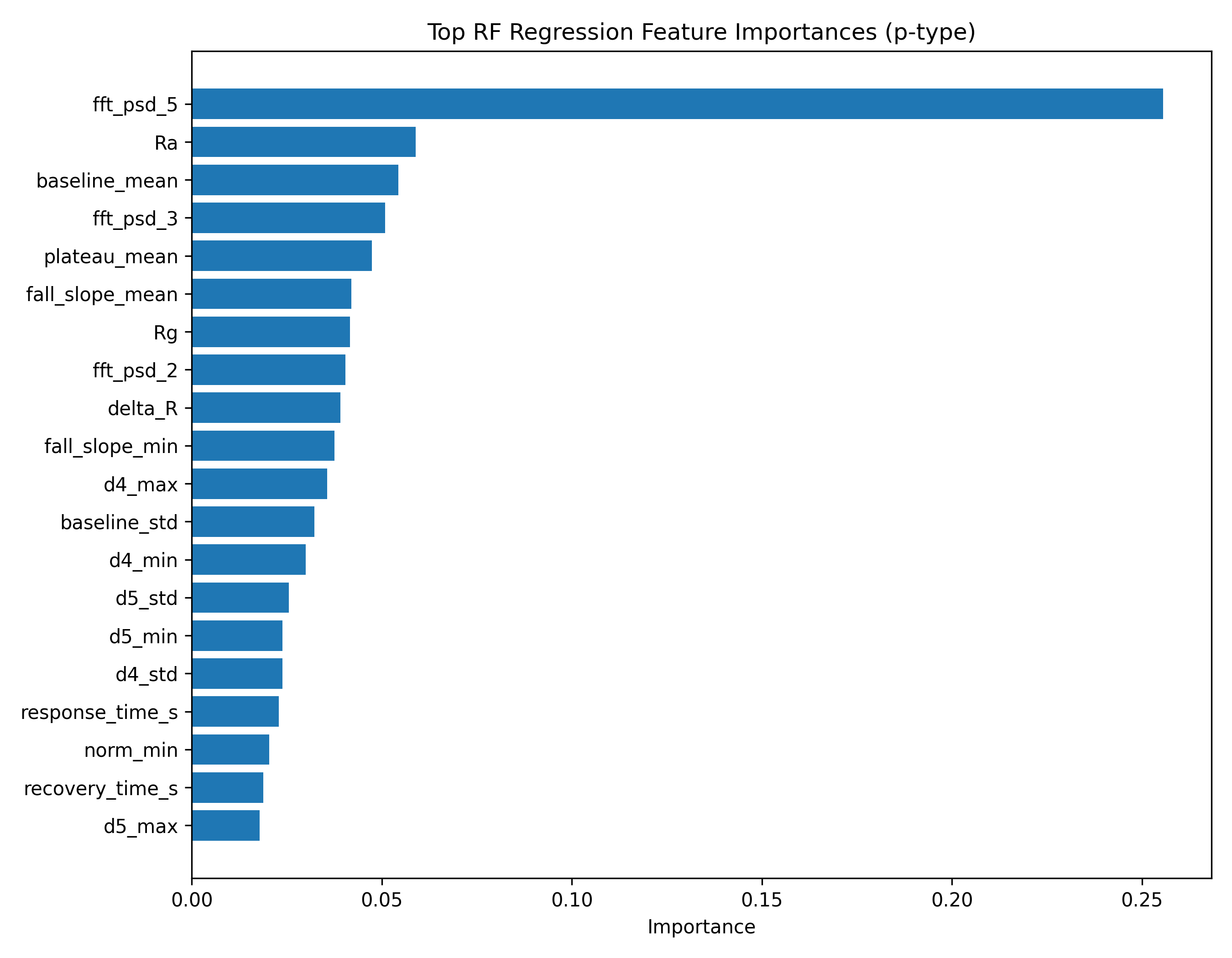}
    \caption{RF (p-type)}
\end{subfigure}
\hfill
\begin{subfigure}[t]{0.245\textwidth}
    \centering
    \includegraphics[width=\linewidth]{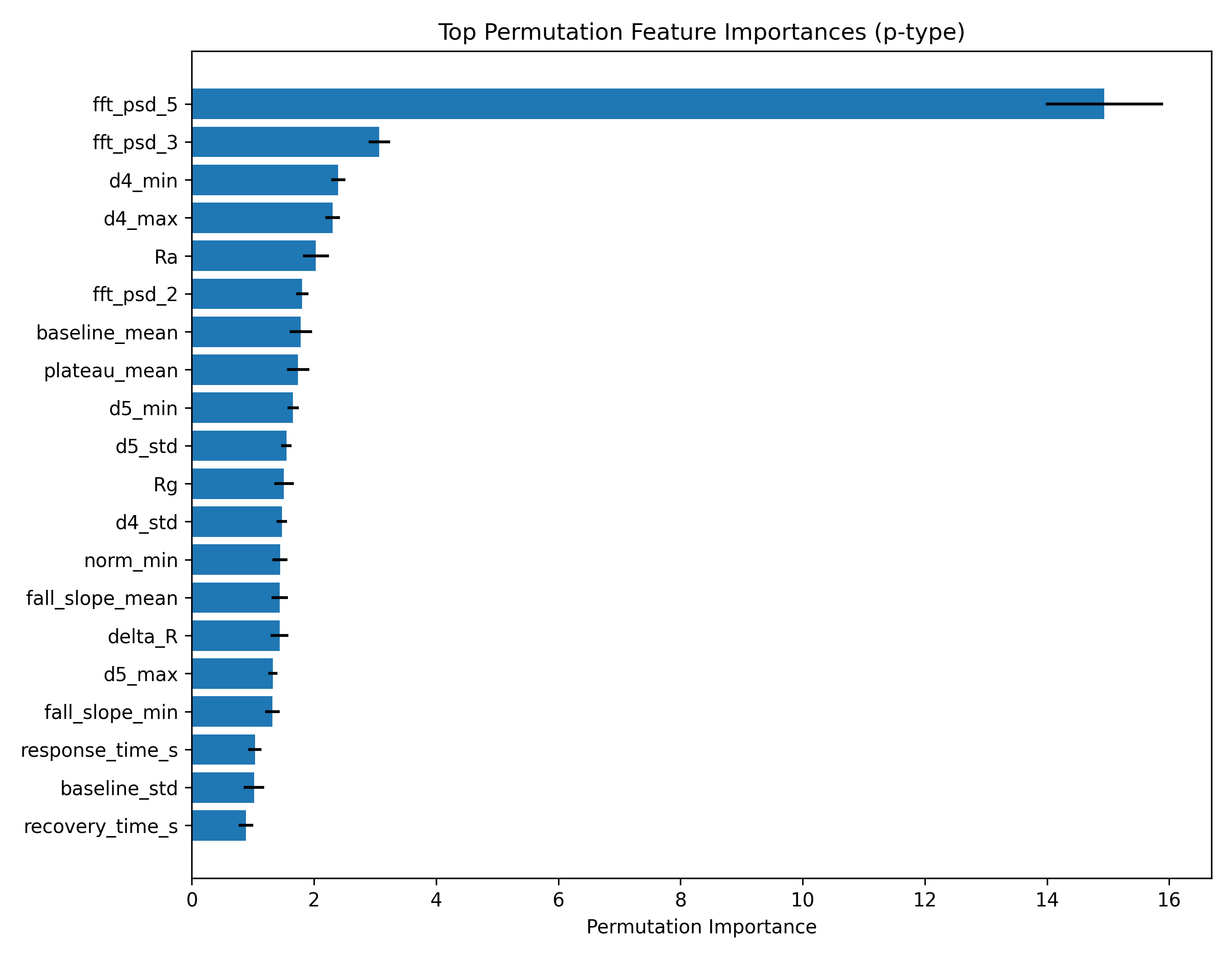}
    \caption{PFI (p-type)}
\end{subfigure}
\hfill
\begin{subfigure}[t]{0.245\textwidth}
    \centering
    \includegraphics[width=\linewidth]{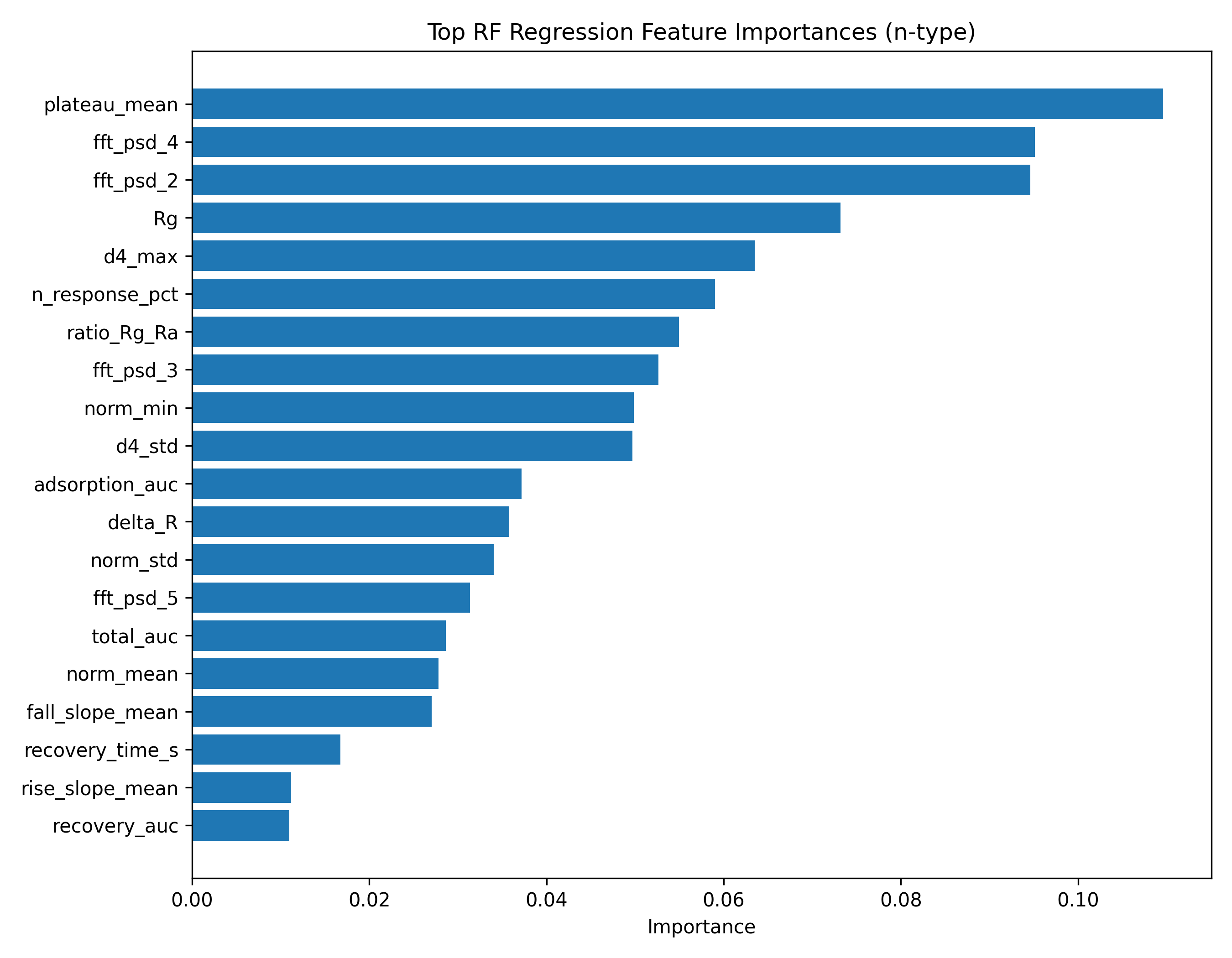}
    \caption{ RF (n-type)}
\end{subfigure}
\hfill
\begin{subfigure}[t]{0.245\textwidth}
    \centering
    \includegraphics[width=\linewidth]{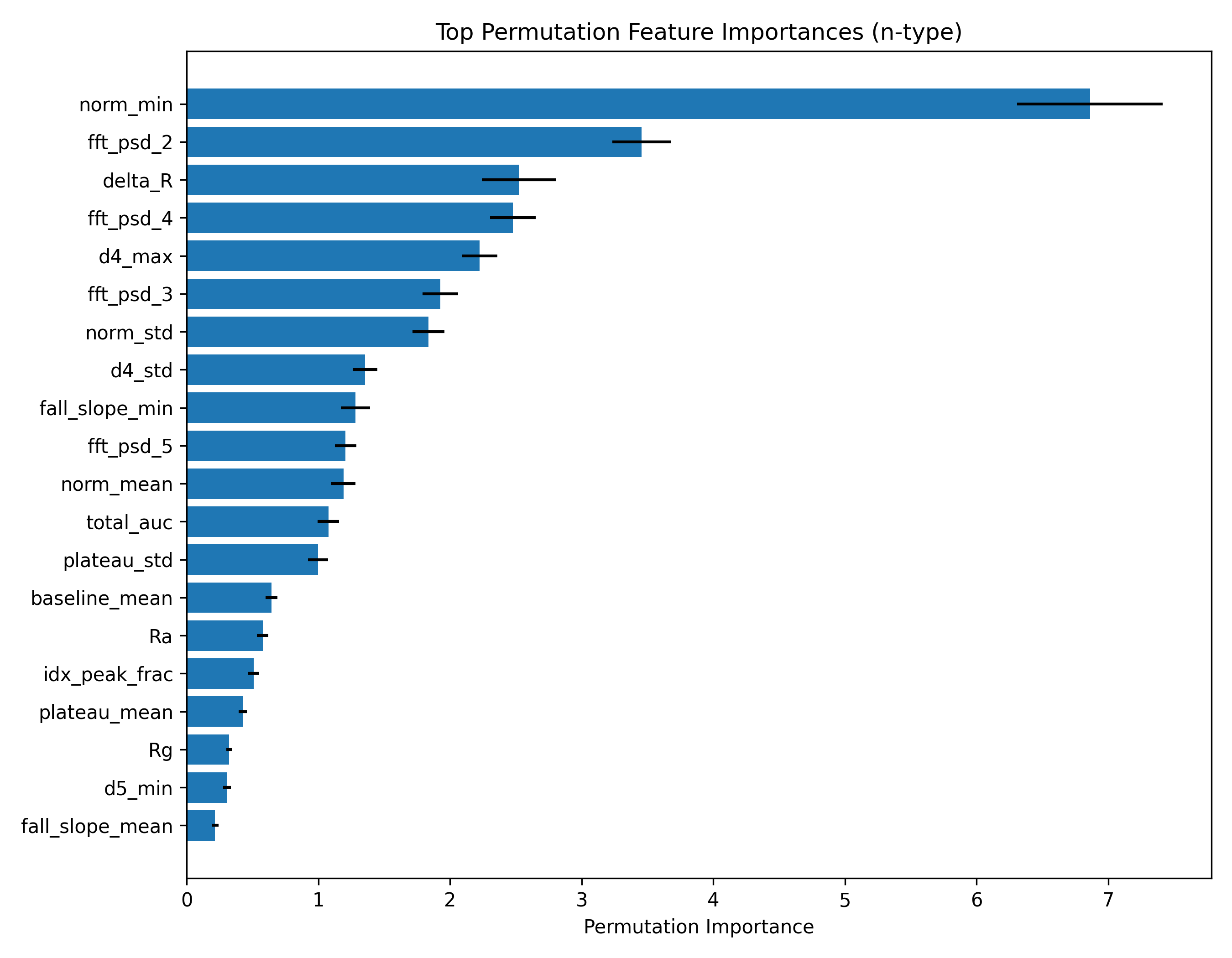}
    \caption{PFI (n-type)}
\end{subfigure}

\caption{Comparison of impurity-based RF and PFI analyses for the FUSED feature representation.}
\label{fig:feature_importance_comparison}
\end{figure*}

Fig.~\ref{fig:feature_importance_comparison} compares the impurity-based RF and PFI analyses for the p-type and n-type sensing regimes. For both sensing regimes, the two methods produce broadly consistent feature rankings, although minor differences are observed in the relative importance of individual features. This agreement confirms the robustness of the identified key predictors and indicates that the predictive success of the proposed framework is driven by physically meaningful sensing characteristics rather than artifacts of a particular feature-importance measure.

% ---------------------------------------------------------
\subsection{Comparative interpretation of the p-type and n-type regimes}
\label{subsec:comparative_interpretation}
% ---------------------------------------------------------

The most important scientific conclusion of the present study is not merely that leakage-aware, physics-guided machine learning performs well on both branches of the mixed-phase SnO-SnO\(_2\) sensing platform. Rather, the p-type and n-type regimes support \emph{complementary modes of concentration inference}. This regime-wise contrast is one of the central findings of the study and provides a physically meaningful synthesis of the grouped classification and regression results.

The p-type branch is the more favorable regime for discrete concentration discrimination. Under
grouped validation, it achieves the strongest classification performance of the study, with the
best fused classifier reaching
\(
\mathrm{Accuracy}=96.50\%,\,
\mathrm{Balanced\ Accuracy}=96.50\%,\,
\mathrm{Macro}\text{-}F_1=96.52\%.
\)
This indicates that the p-type transient responses form sharply separated concentration-dependent
clusters in the engineered feature space, making this regime especially well suited to
concentration-level classification.

By contrast, the n-type branch is the more favorable regime for continuous concentration
quantification. In the grouped regression setting, it achieves the strongest overall regression
result of the study, with the fused Random Forest regressor on the log-target scale yielding
\(
\mathrm{MAE}=1.476\ \mathrm{ppm},\,
\mathrm{RMSE}=2.970\ \mathrm{ppm},\,
\mathrm{MAPE}=24.23\%,\,
R^2=0.9918.
\)
This indicates that the n-type transient manifold supports an especially strong and smoothly
learnable concentration-response relation, even though neighboring concentration levels are less
sharply separated in the discrete classification sense.

These two observations are not contradictory; rather, they suggest that the two conductive regimes
encode concentration information in different ways. The p-type branch appears to generate more
sharply separated class-like transient signatures, whereas the n-type branch appears to produce a
more continuously ordered and highly learnable concentration-response manifold. This is a
scientifically significant result because it shows that the same mixed-phase sensor platform can
support different sensing-intelligence objectives depending on the operative conduction regime.

A second conclusion is methodological and is common to both regimes. Across the grouped
classification and grouped regression studies, the {FUSED} representation consistently
provides the strongest overall performance, while the {PHYSICS} representation remains
highly competitive on its own. This strongly supports the central design philosophy of the paper:
physically interpretable transient descriptors should remain the scientific core of the learning
pipeline, while compact FFT and DWT summaries act as auxiliary refinements rather than as
stand-alone substitutes for physically grounded representation.
% =========================================================
\section{Conclusion}
\label{sec:conclusion}
% =========================================================

This work presented a physics-guided machine-learning framework for carbon monoxide
concentration inference from experimentally measured resistance transients of a mixed-phase
SnO-SnO\(_2\) sensor exhibiting temperature-dependent p-n switching. Rather than treating the
problem as a purely black-box learning task, the proposed methodology was designed as a
computational extension of an experimentally established sensing platform, combining cycle-level
transient analysis, physically interpretable feature construction, and leakage-aware grouped
cross-validation.

The results show that experimentally derived cycle-level resistance transients carry rich concentration information in both sensing regimes. In the p-type branch, the
strongest grouped classification result is obtained by the fused representation with Random Forest,
reaching \(\mathrm{Accuracy}=96.50\%\), \(\mathrm{Balanced\ Accuracy}=96.50\%\), and
\(\mathrm{Macro}\text{-}F_1=96.52\%\), indicating highly discriminative concentration-dependent
transient signatures. In the n-type branch, grouped classification is more moderate, but continuous
concentration regression is exceptionally strong: the best fused Random Forest regressor on the
log-target scale achieves \(\mathrm{MAE}=1.476\ \mathrm{ppm}\), \(\mathrm{RMSE}=2.970\ \mathrm{ppm}\),
\(\mathrm{MAPE}=24.23\%\), and \(R^2=0.9918\).

A central conclusion of the study is therefore regime dependent. The p-type branch is especially
favorable for discrete concentration-level discrimination, whereas the n-type branch is especially
favorable for high-fidelity continuous concentration estimation. This dual-regime behavior is a
meaningful outcome of the mixed-phase SnO-SnO\(_2\) platform and shows that the same sensor can
support different concentration-inference objectives depending on the operative regime.

Across both regimes, the fused representation provides the strongest overall performance, while the
physics-guided descriptor block remains highly competitive on its own. Furthermore, the proposed
feature-based framework consistently outperformed the LSTM baseline, highlighting the effectiveness
of physics-guided feature engineering for the considered sensing data. This supports the main
methodological message of the paper: physically interpretable transient descriptors should remain
the scientific core of intelligent gas-sensing analysis, while compact FFT and DWT summaries are
best viewed as auxiliary refinements rather than substitutes for physically grounded sensing
features. Overall, the present study provides an experimentally grounded, interpretable
machine-learning framework for concentration inference from resistance transients in mixed-phase
SnO-SnO\(_2\) carbon-monoxide sensors.

Finally, this study was validated using CO sensing data acquired from mixed-phase SnO--SnO$_2$ sensors. Although the proposed framework was developed for this sensing configuration, its workflow is not inherently limited to a specific gas or sensing material. Nevertheless, the physics-aware descriptors may require adaptation for different sensing mechanisms. Therefore, future work will focus on validating the proposed framework on other gases and metal-oxide sensors using dedicated experimental datasets. In addition, although preliminary experiments with more complex deep-learning architectures yielded inferior performance on the available dataset, revisiting such models using larger and more diverse datasets remains an interesting direction for future research.

\end{document}